\documentclass[useAMS,usenatbib]{mn2e}

\usepackage{psfig}
\usepackage{graphicx}
\usepackage{times}
\usepackage{natbib}
\usepackage[latin1]{inputenc}

\newif\ifAMStwofonts
\AMStwofontstrue

\newcommand{\bx}{{\bf x}}
\newcommand{\bv}{{\bf v}}
\newcommand{\be}{\begin{equation}}  
\newcommand{\ee}{\end{equation}}
\newcommand{\ba}{\begin{eqnarray}}
\newcommand{\ea}{\end{eqnarray}}
\newcommand{\brr}{\begin{array}}
\newcommand{\err}{\end{array}}
\newcommand{\bc}{\begin{center}}
\newcommand{\ec}{\end{center}}

\newcommand{\lb}{{\left<\right.}}
\newcommand{\rb}{{\left.\right>}}

\newcommand{\vel}{\,{\rm km\,s^{-1}}}

\newcommand{\gadget}{{\footnotesize {\sc GADGET-3~}}}

\newcommand{\GADGET}{{\footnotesize {\sc GADGET}}}
\newcommand{\TradSPH}{{\footnotesize {\sc TradSPH}}}
\newcommand{\GSPH}{{\footnotesize {\sc GSPH}}}
\newcommand{\GSPHCW}{{\footnotesize {\sc GSPH-CW}}}
\newcommand{\GSPHI}{{\footnotesize {\sc GSPH-I02}}}
\newcommand{\GSPHord}{{\footnotesize {\sc GSPH-1ord}}}
\newcommand{\GSPHVLin}{{\footnotesize {\sc GSPH-VLin}}}
\pssilent 

\newcommand{\mincir}{\raise
  -2.truept\hbox{\rlap{\hbox{$\sim$}}\raise5.truept \hbox{$<$}\ }}
\newcommand{\magcir}{\raise
  -2.truept\hbox{\rlap{\hbox{$\sim$}}\raise5.truept \hbox{$>$}\ }}
\newcommand{\siml}{\raise
  -2.truept\hbox{\rlap{\hbox{$\sim$}}\raise5.truept \hbox{$<$}\ }}
\newcommand{\simg}{\raise
  -2.truept\hbox{\rlap{\hbox{$\sim$}}\raise5.truept \hbox{$>$}\ }}

\title[Hydrodynamic simulations with the Godunov SPH] 
  {Hydrodynamic simulations with the Godunov SPH}
  \author[Murante, Borgani, Brunino \& Cha] {G. Murante$^{1}$,
    S. Borgani$^{2,3,4}$, R. Brunino$^5$ \& S.-H. Cha$^6$
  \\~\\
  $^1$ INAF -- Istituto Nazionale di Astrofisica -- Osservatorio
  Astronomico di Torino, Str. Osservatorio 25, I-10025, Pino Torinese,
  Torino, Italy (murante@oato.inaf.it)\\
  $^2$ Astronomy Unit, Department of Physics, University of Trieste, via
  Tiepolo 11, I-34131 Trieste, Italy (borgani@oats.inaf.it)\\
  $^3$ INAF -- Istituto Nazionale di Astrofisica, via Tiepolo 11,
  I-34131 Trieste, Italy\\
  $^4$ INFN -- Istituto Nazionale di Fisica Nucleare, Trieste, Italy\\
  $^5$ CINECA, High Performance System Division, via Magnanelli 6/3,
  I-40033 Casalecchio di Reno, Bologna, Italy (r.brunino@cineca.it)\\
  $^6$ Department of Physics \& Astronomy, University of Leicester,
  Leicester LE1 7RH, UK\\
}

\begin{document}
\label{firstpage}
\maketitle

\begin{abstract}
  We present results based on an implementation of the Godunov
  Smoothed Particle Hydrodynamics (GSPH), originally developed by
  \cite{inutsuka02}, in the GADGET-3 hydrodynamic code. We first review
  the derivation of the GSPH discretization of the equations of moment
  and energy conservation, starting from the convolution of these
  equations with the interpolating kernel. The two most important
  aspects of the numerical implementation of these equations are {\em
    (a)} the appearance of fluid velocity and pressure obtained from
  the solution of the Riemann problem between each pair of particles,
  and {\em (b)} the absence of an artificial viscosity term.  We carry
  out three different controlled hydrodynamical three-dimensional
  tests, namely the Sod shock tube, the development of
  Kelvin-Helmholtz instabilities in a shear flow test, and the
  ``blob'' test describing the evolution of a cold cloud moving
  against a hot wind. 

  The results of our tests confirm and extend in a number of aspects
  those recently obtained by \cite{cha10}: {(i)} GSPH provides a much
  improved description of contact discontinuities, with respect to
  SPH, thus avoiding the appearance of spurious pressure forces;
  {(ii)} GSPH is able to follow the development of gas-dynamical
  instabilities, such as the Kevin--Helmholtz and the Rayleigh-Taylor
  ones; {(iii)} as a result, GSPH describes the development of curl
  structures in the shear-flow test and the dissolution of the cold
  cloud in the ``blob'' test.

  Besides comparing the results of GSPH with those from standard SPH
  implementations, we also discuss in detail the effect on the
  performances of GSPH of changing different aspects of its
  implementation: choice of the number of neighbours, accuracy of the
  interpolation procedure to locate the interface between two fluid
  elements (particles) for the solution of the Riemann problem, order
  of the reconstruction for the assignment of variables at the
  interface, choice of the limiter to prevent oscillations of
  interpolated quantities in the solution of the Riemann Problem. The
  results of our tests demonstrate that GSPH is in fact a highly
  promising hydrodynamic scheme, also to be coupled to an N-body
  solver, for astrophysical and cosmological applications.
\end{abstract}


\section{Introduction}
Lagrangian hydrodynamic methods find a natural field of application in
astrophysics and cosmology. Thanks to their intrinsically adaptive
nature, they are well suited to deal with large ranges of scales and
densities, as well as with the complex geometries of the typical
problems involved in astrophysics. Among such Lagrangian methods,
Smoothed Particle Hydrodynamics \citep[SPH; ][]{gingold77,lucy77} has
been, and currently is, by far the most widely used scheme
\citep[e.g., ][for recent reviews on SPH and on its applications to
astrophysical problems]{monaghan05,rosswog09,springel10b}. The fluid
representation by particles that move with the flow makes SPH
providing a solution of the Euler equation for an inviscid fluid in
Lagrangian coordinates.  Remarkable advantages of this representation
are its intrinsic Galilean invariance and the possibility to easily
couple it to N--body solvers describing the dynamics of
self--gravitating fluids, or to a fluid feeling an external
gravitational potential.

Despite such advantages, a number of intrinsic limitations of SPH have
been recognised. Historically, the first one is related to the
difficulty that SPH has in capturing shocks and contact
discontinuities. At shock fronts, the Rankine-Hugoniot jumping
conditions predict specific entropy
of the fluid to increase through the conversion of mechanical energy
into internal energy, thus implying that an inviscid description of
the fluid can not be valid any longer. To deal with this problem, one
needs to introduce in SPH an artificial viscosity to dissipate local
velocity differences and convert them into heat
\citep[e.g.,][]{monaghan_gingold83,balsara95,monaghan97,lombardi99}. The
typical viscosity required for this is larger than the natural
viscosity of the fluid. This leads to the danger that it may also
provide spurious effects away from shock regions, e.g. causing an
unphysical damping of turbulent motions or spurious angular momentum
transport in differentially rotating discs. Therefore, any scheme of
artificial viscosity need to be tuned so as to be localised as much as
possible at shock regions. Indeed, attempts have been devoted to
design schemes in which artificial viscosity is reduced or even
eliminated away from shocks
\citep[e.g.,][]{morris_monaghan97,dolag05,cullen_dehnen10}. 

Another, possibly more serious limitation, of SPH lies in its
difficulty to correctly describe fluid instabilities and mixing at
the boundaries between different fluid phases. The main reasons for
this are due to the limitations of SPH in computing gradients across
discontinuities and to the intrinsic lack of diffusivity of SPH, which
makes entropy to be conserved within the kernel scale \citep[see][and
references therein, for a detailed discussion]{read10}. This
limitation of SPH has been highlighted by \cite{agertz07}. These
authors carried out a comparison between different SPH and Eulerian
grid codes, with the aim of comparing their relative capability of
describing the onset of fluidodynamical instabilities in specific test
cases. One of the main results of this analysis was that the incorrect
description of contact discontinuities provided by SPH causes a sort
of spurious surface tension to appear, which prevents the development
of Kelvin--Helmholtz (KH) and Rayleigh--Taylor (RT) instabilities. 

Stimulated by this analysis, a number of authors proposed different
approaches to improve SPH performances. \cite{price08} suggested that
discontinuity of entropy at fluid interfaces, combined with a smooth
variation of density, causes the appearance of ``pressure blips'' as a
consequence of a spurious surface tension. To solve this problem, he
introduces a thermal conduction term which acts as a diffusive term in
the energy equation, thus improving the capability of SPH of
describing mixing and the development of instabilities \citep[see
also][]{merlin10}. A similar argument was also presented by
\cite{wadsley08}, who resorted to a sub-grid model of turbulence as
the source of thermal diffusion. \cite{read10} adopted a different
approach. They employed a different kernel, a much larger number
of neighbours, and a modified density estimation formula to control the
errors in the estimates of gradients, so as to obtain a better
representation of mixing of different phases in shear layers.

Well before this vivid debate on the limitations of SPH started,
\cite{inutsuka02} (I02 hereafter) proposed a novel approach to
Lagrangian hydrodynamics. This approach was based on two main
considerations. First, the standard SPH approach is based on assuming
that coarse--grained thermodynamic quantities (i.e. density,
pressure, entropy), assigned at the particle positions by kernel
convolution, can be evolved through the equation of
fluido-dynamics. However, strictly speaking, equation of
fluido-dynamics describe the evolution of micro--physical (i.e. non
coarse-grained) quantities. Therefore, a self--consistent particle
description of fluido-dynamics would require the coarse-graining
procedure to be applied to the equations evolving micro-physical
variables, rather than to the micro-physical variables themselves. The
two operations, i.e. coarse--graining and dynamical evolution, in
general do not commute. It is the violation of this commutation that
originates the approximation in the fluido--dynamical description
provided by SPH. Second, in deriving the implementation, through
particle description, of the equations of evolution, I02 devoted
special attention to keep the order of spatial accuracy (i.e., in
kernel smoothing length) fixed to second-order. The natural way of
implementing this SPH formulation is by computing the exchange of
hydrodynamic forces and momenta between pairs of particles by using a
Riemann solver, analogous to the grid-based second-order Godunov
scheme \citep{vanleer79}. Therefore, the resulting Godunov Smoothed Particle
Hydrodynamics (GSPH hereafter) method has the extra benefit of
not requiring the introduction of an artificial viscosity term, since shocks
are now naturally described as solutions of the Riemann problem (RP
hereafter). I02 showed with one-dimensional tests that the mixing
associated to the energy and momentum exchange naturally prevents the
formation of ``pressure blips'' at contact discontinuities \citep[see
also][]{cha_whitworth03}. More recently, \cite{cha10} showed that GSPH
also provides a much improved description of the development of KH
instabilities through two--dimensional tests. 

In this paper we present results based on our implementation of GSPH
on the GADGET-3 simulation code \citep{springel05}. Using standard
three--dimensional hydrodynamic tests, we aim at demonstrating the
capability of GSPH of describing mixing and development of
fluidodynamical instabilities at interfaces. In our analysis we will
pay special attention to {\em i)} highlight the fundamental
differences with respect to the standard SPH approach; {\em ii)}
discuss the effect of changing relevant aspects of the implementation
of the Riemann solver. 

We note that \cite{springel10a} recently proposed a novel
particle--based method, in which particle positions are used to
construct an unstructured moving mesh, according to the Voronoi
tessellation. Similarly to the GSPH, also in this case fluid equations
are solved with the finite volume Eulerian Godunov scheme, with fluxes
across the boundaries of the Voronoi poliedra provided by the solution
of the Riemann problem between particle pairs. While this scheme and
GSPH appear to be similar in spirit, there are a number of fundamental
differences. The most important is probably represented by the fact
that, while GSPH always refers to the volumes as defined by the kernel
smoothing length, the scheme proposed by \cite{springel10a} uses the
partitioning of the computational domain given by the Voronoi
tessellation.

The scheme of our paper is as follows. Section 2 gives the basis of
our GSPH formulation. In Section 3 we describe the details of our
implementation of GSPH in GADGET. In Section 4 we describe our results
for the hydrodynamical test of the shock tube, the shear flows and of
the ``blob'' test. In Section 5 we draw our conclusions. The Appendix
contains the expression for interpolating volumes and for the position
of the interface.

\section{Basics of the Godunov SPH}
In the following we provide a short description of the approach at the
basis of the Godunov SPH (GSPH) method, while we refer to the paper by
\cite{inutsuka02} for a more complete formal derivation.

Let us introduce the convolution of a physical function $f(\bx)$ with
the kernel function,
\be
\lb f\rb(\bx)\,\equiv \,\int f(\bx')W(\bx-\bx';h)\,d\bx'\,,
\label{eq:fc}
\ee 
where, as usual, $h$ denotes the kernel size, that for the moment we
assume to be spatially constant, and the integral is
performed over the whole spatial domain.  Starting from the above
expression, it is easy to demonstrate that $\lb\nabla f\rb =\nabla \lb
f \rb$.

Defining the density field at the position $\bx$ as given by the
sum of the kernel contributions at particles positions $\bx_j$,
$\rho(\bx)=\sum_jm_j \,W(\bx-\bx_j;h)$, we have then the identities
\be
1=\sum_j{m_j\over \rho(\bx)}\,W(\bx-\bx';h)~~~~;~~~~
0=\sum_j m_j\,\nabla \left[{W(\bx-\bx_j;h) \over \rho(\bx)}\right]\,.
\label{eq:iden}
\ee
Using the first of the two above identities, the expression of the
kernel convolution can be cast in the form
\begin{equation}
f_i\equiv\lb f\rb (\bx_i)=\int \sum_j m_j\,{f(\bx')\over
  \rho(\bx')} W(\bx'-\bx_i;h) W(\bx'-\bx_j;h)d\bx'.
\end{equation}
We note that the expression for the kernel convolution in
the standard SPH approach is recovered under the approximation
$W(\bx-\bx';h)=\delta_D(\bx-\bx')$.

To derive the equation of evolution for particles, we start from the
kernel convolution of the equation of motion,
\be
\int {d\bv(\bx)\over dt}\,W(\bx-\bx';h)\,d\bx =-\int{\nabla
  P(\bx)\over \rho(x)}\,W(\bx-\bx';h)\,d\bx\,,
\label{eq:eom_con}
\ee
where $\bv(\bx)$ is the velocity field and $P(\bx)$ the pressure field
of the fluid. 
Since $\ddot{\bx}_i\equiv \int {d\bv(\bx)\over
  dt}\,W(\bx-\bx';h)\,d\bx$ describes the motion of the $i$-th particle
position, integrating by part the {\em r.h.s.} of the above equation
and using the first identity in Eq.(\ref{eq:iden}),  we
obtain the following expression for the equation of motion:
\be
m_i\ddot{\bx}_i\,=\,-m_i\sum_j m_j\int{P(\bx)\over
  \rho(\bx)^2}\,\left[\partial_i
  -\partial_j\right]W_i W_j\,d\bx\,.
\label{eq:eom} 
\ee
where we introduced the notations $\partial_i=\partial/\partial
\bx_i$ and $W_i=W(\bx-\bx_i;h)$. 

As for the energy equation, the coarse--grained representation for the
evolution of the specific internal energy $u(\bx)$ can be
written as
\ba
&&\int{du(\bx)\over dt}\,W(\bx-\bx';h)\,d\bx= \nonumber \\
&&-\int{P(\bx)\over
  \rho(\bx)}\left[\nabla\cdot \bv\right]W(\bx-\bx';h)\,d\bx\,.
\label{eq:eneq_con}
\ea
After rearranging the {\em r.h.s.} of the above equation and using the
approximation 
\ba
&&\int {\bv \cdot \nabla P(\bx)\over
  \rho(\bx)}\,W(|\bx-\bx'|;h)\,d\bx=\nonumber \\
&&\int {\dot \bx_i \cdot \nabla P(\bx)\over
  \rho(\bx)}\,W(|\bx-\bx'|;h)\,d\bx + O(h^2)\,,
\ea
the equation for the evolution of the internal energy of the $i$-th
particle can be written as 

\be
\dot u_i = \sum_j m_j \int \frac{P(x)}{\rho^2(x)}(v-\dot x_i) \cdot
(\partial_i - \partial_j)W_i W_j
\label{eq:eneq}
\ee

We point out that Eqs.(\ref{eq:eom}) and (\ref{eq:eneq}) replace in
the GPSH approach the corresponding equations of motion and energy of
the standard SPH approach. As demonstrated by \cite{inutsuka02}, they
provide a description of the
coarse--grained equations of fluido-dynamics of
Eqs. (\ref{eq:eom_con}) and (\ref{eq:eneq_con}), which is second-order
accurate in spatial resolution (i.e. in kernel smoothing length). 

\section{Implementation}
\label{s:impl}
In this section we describe the numerical implementation of the GSPH
equations of evolution. We remind that we implemented these equations
in the massively-parallel GADGET-3 simulation code. GADGET-3 is a
N-body/hydrodynamic code in which entropy-conserving SPH
\citep{springel_hernquist02} is coupled to a TreePM N--body solver to
describe gravity. The code has fully adaptive time-stepping. Domain
decomposition is carried out by using a space-filling Peano-Hilbert
curve, which is split into segments assigned to different computing
units. In this respect, GADGET-3 represents a substantial improvement
with respect to the previous GADGET-2 version \cite{springel05}, in
that disjointed segments of the Peano-Hilbert curve can be assigned to
a single computing unit, so as to achieve an optimal work-load
balance. Besides the reference entropy conserving SPH formulation,
GADGET also includes a switch to standard energy conserving SPH. Our
implementation of GSPH replaces the equations of standard SPH.

\subsection{The GSPH equations}
I02 described how to evaluate the spatial integrals appearing in
Eqs.(\ref{eq:eom}) and (\ref{eq:eneq}), under the assumption of
Gaussian kernel,
\be
W(x,h)\,=\,(\pi h^2)^{-d/2} e^{-x^2/h^2}\,.
\label{eq:gauk}
\ee
where $d$ is the number of dimensions.
To this purpose, for a given pair of particles having coordinates
$\bx_i$ and $\bx_j$, one
defines the $s$-axis, which is parallel to the direction of the
separation vector $\bx_i-\bx_j$, with origin
at $(\bx_i-\bx_j)/2$. We also denote with $s_i$ and $s_j$ the
components of the $\bx_i$ and $\bx_j$ vectors along the $s$-axis, so
that $\Delta s_{ij}=s_i-s_j=|\bx_i-\bx_j|$. Defining then the specific
volume occupied by a fluid element of density $\rho(\bx)$ as
$V(\bx)=1/\rho(\bx)$, then its gradient is $\nabla V(\bx)=-\sum_j
m_j\nabla(\bx-\bx_i;h)/\rho^2(\bx)$. It is then possible to
demonstrate that after expanding $1/\rho^2$ to 
the first order in the direction parallel
to the vector $\bx_i-\bx_j$, the equations of
evolution can be rewritten in the form
\be
{\Delta \dot \bx_i\over \Delta t}=-2\sum_j m_j P^*
V_{i,j}^2(h)\partial_i W(\bx_i-\bx_j;\sqrt 
2h)\,,
\label{eq:eom_gsph}
\ee 
\be
{\Delta u_i\over \Delta t}=-2\sum_j m_j P^*[\bv^*-\dot
\bx_i]V_{i,j}^2(h) \partial_i W(\bx_i-\bx_j;\sqrt 2h)\,.
\label{eq:eneq_gsph}
\ee 
In the above equations $\Delta$ indicates finite difference of
each variable, $\bx_i$ is the time-centred velocity of the $i$-th
particle, while $P^*$ and $\bv^*$ are provided by the solution of the
Riemann problem between the $i$-th and the $j$-th particles.  The
  use of $P^*$ instead of $P(\bx)$ is justified using a linear
interpolation 
for $P(\bx)$, and evaluating it at the position
  $s^*_{i,j}$ (See the Appendix for details).
Furthermore, the
quantity $V_{i,j}^2$ accounts for the expansion of the $1/\rho^2$ term
and can be expressed through the kernel convolution according to the
relation \ba &&\int
\rho^{-2}(\bx)\,W(\bx-\bx_i)\,W(\bx-\bx_j)\,d\bx=\nonumber \\
&&V_{i,j}^2(h)\,W(\bx_i-\bx_j;\sqrt 2h)\,.
\label{eq:vij} 
\ea 
We point out that the possibility of factoring out in the above
equation the dependence on the separation vector of the particle pair,
and the $\sqrt 2$ factor appearing in front of the smoothing length in
the {\em r.h.s.} stem from the assumption of Gaussian kernel.

We provide in Appendix the expressions for the position
of the interface and for the $V_{i,j}$ quantities in the case of
linear and of cubic spline interpolation for the $V(\bx)$ in the
$s$-coordinate. In the following we will use the more accurate cubic
spline in our reference \GSPH\ formulation. We will also show the
effect of using a linear interpolation for volumes, in the shear flow
and blob tests.

We also point out that the derivation of the \GSPH\ equations of evolution
(\ref{eq:eom_gsph}) and (\ref{eq:eneq_gsph}) have been derived by
assuming a constant value for the kernel smoothing length $h$. In the
case of adaptive smoothing, the formal derivation that leads to
Eqs.(\ref{eq:eom}) and (\ref{eq:eneq}) can be repeated by just
replacing $h$ with $h(\bx)$. However, for a spatially varying
smoothing length the convolution integrals leading to the \GSPH\
equations (\ref{eq:eom_gsph}) and (\ref{eq:eneq_gsph}) can not be done
analytically. In this case, I02 made the ansatz that $h_i=h(\bx_i)$
has to be used for half of the integration volume and $h_j=h(\bx_j)$
for the other half. This leads to the following expressions for the \GSPH\
evolution equations in the presence of adaptive smoothing:
\ba
{\Delta \dot \bx_i\over \Delta t}&=&\sum_j m_j P^*
\biggl[V_{i,j}^2(h_i)\partial_i W(\bx_i-\bx_j;\sqrt 2h_i)\nonumber \\
&+&V_{i,j}^2(h_j)\partial_i W(\bx_i-\bx_j;\sqrt 2h_j)\biggr]
\,,
\label{eq:eom_h}
\ea
\ba
{\Delta u_i\over \Delta t}&=&-\sum_j m_j P^*[\bv^*-\dot \bx_i]
\biggl[V_{i,j}^2(h_i) \partial_i W(\bx_i-\bx_j;\sqrt 2h_i)\nonumber \\
&+&V_{i,j}^2(h_j) \partial_i W(\bx_i-\bx_j;\sqrt 2h_j)\biggr]\,.
\label{eq:eneq_h}
\ea

In the following we will estimate the local value of $h(\bx)$ by assuming
that the kernel contains a fixed number $N_{\rm neigh}$ of particles. 
The Gaussian kernel has not a compact support, thus implying that each
particle should in principle interact with all the other
particles. This would result in an extremely expensive calculation,
with a time cost $t_{\rm CPU} \propto N_{\rm part}^2$, scaling with the
square of the particle number. To avoid this, we simply truncate the
Gaussian kernel at a distance $r = 3 h$, the neglected contribution
of the kernel being of the order of $10^{-5}$.

We also discuss the
effect of varying the number of neighbours in the case of the KH test
(see Section \ref{sec:KH} below).  We note that this criterion to
choose the number of neighbours is different from that adopted by I02,
which is instead based on the requirement that the resulting $h(\bx)$
is not varying much within the neighbourhood of each particle.

Furthermore, for each particle $i$ the sums appearing in
Eqs.~(\ref{eq:eom_h}) and (\ref{eq:eneq_h}) must be performed over all
neighbours within a distance $\sqrt 2 \,h_i$ or $\sqrt 2 \,h_j$. This
further increases the number of neighbours thus increasing the
computational cost.

As discussed by \cite{cha10}, the standard SPH evaluation of gas
density can generate an unphysical repulsive force in some particular
particle distributions, for example a non uniform one. This problem is
prevented by defining density as an even function for the exchange of
particle positions. Therefore, we symmetrize our density estimate with
respect to the Gaussian kernel, according to 
\ba 
\rho(\bx_i) =
\sum_{|\bx_j-\bx_i|<\max\{h_i,h_j\}} m_j W_{ij}\,,
\label{eq:dens_symm}
\ea

where $W_{ij}=[\left[W(\bx_j-\bx_i;
    \sqrt{2}h_i)+W(\bx_i-\bx_j; \sqrt{2}h_j)\right]/2$.
We use this estimate of the gas density in all of our GSPH
formulations, unless otherwise specified. 

We implement Eqs.~(\ref{eq:eom_h}) and (\ref{eq:eneq_h}) in the
GADGET-3 code. Besides avoiding the need of introducing the artificial
viscosity term in the equation of motion, another fundamental
difference between the \GSPH\ and the standard SPH evolution equations
lies in the fact that velocity and pressure terms associated to each
pair of particles are replaced by the corresponding RP solutions,
evaluated at the interface position. This causes the appearance in the
energy equation of the term between square brackets in the {\em
  r.h.s.}, which effectively represents a mixing term for internal
energy. This is inherently different from SPH, which instead provides
a strictly non-diffusive description of the evolution of internal
energy.

However, it is worth pointing out that the above \GSPH\ evolution equations
{\em can not} be obtained by simply replacing pressure and velocity
terms in the standard SPH formulation, with the values provided by the
solution of the Riemann Problem between $i$-th and $j$-th particle. A
further fundamental point of difference lies in the interpolating
volumes $V_{i,j}$. These terms account for the fact that \GSPH\
equations are directly derived from the convolution of the energy and
momentum equations. Indeed \cite{cha_whitworth03} introduced a variant
of these equations, which neglects the convolution integrals and,
therefore, are equivalent to replacing the interpolating volumes with
the values of $1/\rho^2$ computed at the positions of the $i$-th and
$j$-th particle: 
\ba {\Delta \dot \bx_i\over \Delta t}&=&\sum_j m_j P^*
\biggl[{1\over \rho^2(\bx_i)}\partial_i W(\bx_i-\bx_j; h_i)\nonumber \\ 
&+&{1\over \rho^2(\bx_j)}\partial_i W(\bx_i-\bx_j; h_j)\biggr]
\,,
\label{eq:eom_gph}
\ea
\ba
{\Delta u_i\over \Delta t}&=&-\sum_j m_j P^*[\bv^*-\dot \bx_i]
\biggl[{1\over \rho^2(\bx_i)} \partial_i W(\bx_i-\bx_j;
h_i)\nonumber \\ 
&+&{1\over \rho^2(\bx_j)} \partial_i W(\bx_i-\bx_j; h_j)\biggr]\,.
\label{eq:eneq_gph}
\ea 
Since in this formulation of GSPH there is no need to perform any
convolution, we implemented Eqs. \ref{eq:eom_gph} and
\ref{eq:eneq_gph} in the GADGET-3 code using its original B-spline
kernel. Note that the absence of the factor $\sqrt 2$ in front of the
kernel smoothing length $h$ is due to the fact that the above
equations have not been derived from the convolution of two Gaussian
kernels, as in the correct \GSPH\ formulation. Furthermore, following
\cite{cha_whitworth03}, we use for this formulation the standard SPH
computation of gas density, and not the symmetrized one provided by
Eq. \ref{eq:dens_symm}. As we shall demonstrate below with
hydrodynamical tests, the formulation provided by
Eqs. (\ref{eq:eom_gph}) and (\ref{eq:eneq_gph}) turns out to be
exceedingly diffusive and provides an incorrect description of the
development of gas-dynamical instabilities. This highlights the
relevance of properly describing the volume convolution in the
particle description of the equations of fluido-dynamics.

\subsection{The Riemann solver and the slope limiter in GSPH}
We need to know the values of $P_*$ and $v_*$ at the interface
position, for each pair of particles $(i,j)$, in order to use them in
the \GSPH\ Equations (\ref{eq:eom_h}) and (\ref{eq:eneq_h}). In
grid-based hydrodynamical codes, the Godunov method is based on
solving a Riemann Problem (RP) at each cell interface to evaluate
numerical fluxes, which are then used to update the value of
thermodynamical quantities in the cell. In the case of \GSPH, we do not
need to compute fluxes, since we want to preserve the Lagrangian
nature of the hydrodynamic description, while we only need the values
of the (post-shock) velocity $v_*$ and pressure $P_*$ at the
interface.

To solve the RP, we first have to define the right and left
(pre-shock) states ($P_{R,L}$, $v_{R,L}$ and $\rho_{R,L}$). The
simplest choice is to use the SPH values of $P$, $v$ and $\rho$,
associating the $i$-th particle to the {\it right} state and the
$j$-th particle to the {\it left} one. This corresponds in the
original Godunov scheme to a first-order spatial accuracy. Since we
are not interested in computing fluxes, the exact position of the
interface in this first-order scheme is not important, given that
thermodynamical quantities are assumed to be spatially constant for
each state. This would make the first-order scheme in principle well 
suited for our purpose. However, this scheme is known to be highly
dissipative, thus making it not ideal to capture the development of
fluido-dynamical instabilities.

A second-order spatial accuracy can be achieved using a piecewise
linear distribution of thermodynamical quantities. In this case, the
left and right states are defined by the values of ($P$, $v$, $\rho$)
at the interface position $s_*$. The (linear) interpolation of these
variables is obtained using their derivatives along the $s$-axis
\citep[see][for details]{inutsuka02}. We describe in the Appendix how
the position $s_*$ of the interface can be computed in the case of
linear and cubic interpolation of the volume $V(s)$.

\begin{figure}
\includegraphics[scale=0.7]{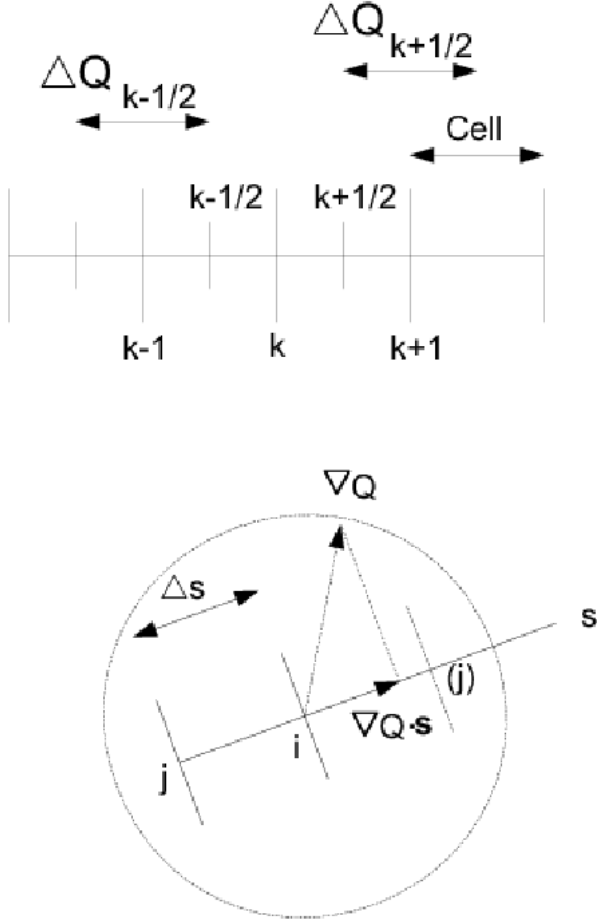}
\caption{Cartoon illustration of how the gradient of a thermodynamical
  quantity $Q$ is computed for the purpose of implementing the limiter
  introduced by \protect\cite{vanleer79}. The upper figure describes
  the standard case of the computation in case the variable $Q$ is
  assigned on a regular grid. The lower figure is for the case,
  relevant for our implementation of \protect\GSPH, in which the
  variable $Q$ is assigned at the positions of the $i$-th and $j$-th
  particles.}
\label{fi:limi}
\end{figure}

In higher-order (i.e. second-order and above) schemes, thermodynamical
quantities must be limited to obtain a stable description of the
discontinuity. This is obtained by implementing a limiter, which is
defined as ``a non-linear algorithm that reduces the high-derivative
content of a subgrid interpolant in order to make it non-oscillatory''
\citep{vanleer06}. For instance, \cite{inutsuka02} imposed that a
second-order reconstruction is performed when the components of
velocity gradients, at the position of the two particles, parallel to
the $s$-axis have the same sign, while one resorts to a first--order
reconstruction in case of discordant signs.

We note that \cite{Godunov59} proved a theorem which states that
  any advection scheme preserving the monotonicity of the solution is
  at most first-order accurate. This holds only if the discretization
  of the advection equation is linear. Thus, non-linear schemes
  are needed to achieve higher order accuracy. On the other hand,
  for them to be useful, higher-order schemes need to include in the
  interpolator a prescription to limit spurious oscillations. For
  example, 
in the context of Eulerian schemes, 
Van Leer (1979) proposed to employ a ``harmonic gradient averaging''
technique, in which the gradient of the thermodynamical quantity $Q$
is the harmonic average of the gradients in the $k+1/2$ and $k+3/2$ cells,
namely:
\be
\Delta Q_{k+1/2} = {{2 \Delta_k Q \Delta_{k+1} Q} \over {\Delta_kQ + \Delta_{k+1}Q}}\,,
\label{eq:vllim0}
\ee
where $\Delta_kQ = Q_{k+1/2}-Q_{k-1/2}$ and we assumed unity value for
the cell size. In this notation, $Q_{k+1/2}$ indicates the
mass-averaged value of a given thermodynamical variable in the cell
$k+1/2$, having boundaries at $k$ and $k+1$, thus with the obvious
extension for the meaning of $Q_{k-1/2}$.  The mass-averaged gradients
of the variable $Q$ at half-cell positions, $\Delta Q_{k\pm1/2}$, are then
used to estimate the right-state and left-state values of $Q$ at the
interface:
\begin{equation}
\begin{array}{lcl}
\vspace{5mm}
Q_{L,R} \!=\!
\!\left\{ \!\begin{array}{l}
\vspace{5mm}
Q_{k\pm 1/2} \mp \Delta Q_{k \pm 1/2}/2 \;\;\mbox{\rm if } \Delta Q_{k-1/2}\Delta Q_{k+1/2}>0,\\
Q_{k\pm 1/2}  \;\;\mbox{\rm if } \Delta Q_{k-1/2}\Delta Q_{k+1/2}<0,
           \end{array}
    \right.
\end{array}
\end{equation}

It can be shown \citep{vanleer06} that such an interpolator
  corresponds to a standard central difference of the quantities $Q$,
  limited by a term of order $(\Delta s)^2$ which depends on the rate
  of change of the quantity itself through its second-order
  derivative. Thus, the harmonic gradient averaging is a good example
  of a non linear interpolator with a built-in limiter. 

In the case of an Eulerian scheme implemented on a regular Cartesian
grid, Eq. \ref{eq:vllim0} uses three adjacent cells, namely $k-1/2$, $k+1/2$
and $k+3/2$. Clearly, this prescription to compute gradients can not be
directly generalised to our GSPH scheme, where the values of the
variable $Q$ are assigned at the positions of the $i$-th and $j$-th
particles, while there is no a third particle to form a regular grid
along the direction of the separation vector $\bx_i-\bx_j$.

In the upper part of Figure \ref{fi:limi} we schematically show how
quantities are evaluated in the Eulerian scheme, while the lower part
shows our extension of the implementation of the \cite{vanleer79}
limiter in the case of the unstructured grid, which is relevant for
the GSPH. In this case, we only have quantities $Q_i$ and $Q_j$
evaluated at the particles positions. We define the equivalent of the
Eulerian $\Delta Q_{k-1/2}$ as:
\be
\Delta Q_1 = {{Q_j - Q_i} \over {\Delta s}}\,,
\label{eq:q1}
\ee where $\Delta s$ it the distance between the two particles. To
evaluate the equivalent of $\Delta Q_{k+1/2}$, we use a ``ghost''
particle $(j)$, which is located the same distance $\Delta s$ from the
$i$-th particle, but in the opposite direction along the $s$
axis. Since we have the SPH estimate of the gradient $(\nabla Q)_i$ at
the position of the $i$-th particle, we can compute the expected value
$Q_{(j)}$ at the position of the ``ghost'' particle as
\be
 Q_{(j)}=Q_i + (\nabla Q)_i \cdot {\bf s} \Delta s\,.
\label{eq:qgh}
\ee
Thus, the equivalent of the quantity $\Delta Q_{k+1/2}$, that we
defined in the case of the regular grid, is
\be
\Delta Q_2= {Q_i + (\nabla Q)_i \cdot {\bf s} \cdot \Delta s -Q_i \over \Delta s} =
  \nabla Q \cdot {\bf s}
\label{eq:q2}
\ee Therefore, $\Delta Q_2$ is simply the projection of the gradient of the
quantity $Q$ on the $s$-axis, In general, it is $\Delta Q_1 \ne \Delta Q_2$, since
$Q_2$ depends on the 3D properties of the field $Q$ through its
gradient computed at the $i$-th particle, while the latter only
depends on the values of the field at the positions of the $i$-th and
$j$-th particles. Of course, the same calculation can be repeated for
the $j$-th particle using a ``ghost'' particle $(i)$, which is defined
in the same way. Therefore, the extension of Eq.(\ref{eq:vllim0}) to
the case of an unstructured grid, defined by particle positions, reads
\begin{equation}
\begin{array}{lcl}
\vspace{5mm}
{{\partial
    Q }\over {\partial s}} \!=\!
\!\left\{ \!\begin{array}{l}
\vspace{5mm}
{{2 \Delta Q_1 \Delta Q_2 } \over { \Delta Q_1 + \Delta Q_2}} \;\;\mbox{\rm if } \Delta Q_1 \Delta Q_2>0,\\
0  \;\;\mbox{\rm if } \Delta Q_1 \Delta Q_2<0,
           \end{array}
    \right.
\end{array}
\end{equation}

Having computed the values of the gradients ${{\partial Q}\over
  {\partial s}}$ at the positions of the $i$-th and $j$-th particles,
we can assign the values of thermodynamical quantities on the left
and right states at the interface, following the same procedure as in
I02. We will refer to this implementation as
``second order reconstruction with van Leer limiter'' \citep[see][ for
a detailed description]{vanleer06} .

Once we have thermodynamical variables assigned on the left and right
states ($P_{R,L}$, $v_{R,L}$, $\rho_{R,L}$), we still have to solve
the RP. Many Riemann solvers do exist in the literature
\citep[e.g.][and references therein]{Toro99}. Here, we use the
iterative solver originally proposed by \citep{vanleer79} and
described in detail by \cite{cha_whitworth03}. We refer to this paper
for a complete description of the implementation of this solver. The
general idea is {\em (i)} to define the Lagrangian shock speed $W$
starting from an initial guess based on the values of $P_*$ and $v_*$;
{\em (ii)} calculate the values of the tangential slopes $Z={dP_*\over
  dv_*}$; {\em(iii)} use these slopes to evaluate the new values of
$P_*$ and $v_*$; {\em (iv)} iterate until the variation in $P_*$ with
respect to the previous iteration falls below a given threshold value,
namely, is less than 1.5 \%.  We also checked that using a different
solver, namely a Newton iterative solver \citep[e.g.][]{vanleer06},
the results of our hydrodynamical tests do not appreciably change.
Both the Van Leer and the Newton solvers are exact. Usually, they
provide a converged result in 5-7 iterations. A significant speed-up
of the code could be obtained by using instead an approximate
one-iteration solver, such as the Harten--Lax--van Leer--Contact
(HLLC) solver proposed by \citep{toro_etal94}.

In summary, the implementation of the Godunov method in our \GSPH\
version of the GADGET-3 code requires the following steps.

\begin{description}
\item[1.] Estimates of the volume function $V(s)$ and of the position
  $s_*$ of the interface between each pair of particles. In our
  implementation, this can be done through either a linear
  interpolation or a cubic spline;
\item[2.] Choice of the spatial order of the reconstruction of the
  thermodynamical values at the interface position. Currently, we have
  implemented first and second order reconstructions. A third order
  scheme, such as the Piece-wise Parabolic Method (PPM), could in
  principle be implemented.
\item[3.] Choice of the limiter, in the case of second-order
  reconstruction. We have implemented both the ``standard''
  reconstruction, as in \cite{inutsuka02}, and the reconstruction
  scheme proposed by \cite{vanleer06}.
\item[4.] Choice of the solver for the RP. We have implemented two
  equivalent exact solvers, namely the iterative solver proposed by
  \cite{vanleer79} and the Newton solver. We will present results
  based on the former solver, while we have verified that identical
  results are obtained by using the Newton solver. 
\end{description}

We provide in Table \ref{t:cod} a description of the variants of the
hydrodynamical schemes that we compare through the tests described in
the following section.

The first one is the original GADGET-3 entropy conserving scheme,
tagged \GADGET. We then use a traditional, energy conserving SPH
scheme (\TradSPH). In this case, however, we use a Gaussian kernel,
rather than the B--spline one implemented in \GADGET. We tag as \GSPH\
our new reference implementation, which uses cubic spline
interpolation for volumes, second-order reconstruction of
thermodynamical variables at the interface, with the limiter and the
iterative solver for the Riemann problem, both proposed by
\cite{vanleer79}. Furthermore, we tag as \GSPHI\ the version based
instead on using the limiter adopted in I02, with \GSPHord\ the
version based on a first-order reconstruction of the thermodynamical
quantities at the RP interface, and with \GSPHVLin\ the version based
on the linear interpolation for the volume function $V(s)$ (see
Appendix). Finally, we also implemented the version of the Godunov
SPH scheme \citep{cha_whitworth03}, which is described by
Eqs.(\ref{eq:eom_gph}) and (\ref{eq:eneq_gph}), rather than by actual
equations of GSPH involving the volume integrals, as in
Eqs.(\ref{eq:eom}) and (\ref{eq:eneq}). We will refer to this scheme
as \GSPHCW.

\begin{table*}
\label{t:cod}
\begin{center}
\begin{tabular}{lclccc}
\hline
 Name & Hydrodynamic scheme & Implementation details & Sod & KH & Blob \\  
\hline
\hline
\\
\GADGET\  & SPH & Entropy conserving with B-spline kernel. & 100 & 442
& 50 \\
\TradSPH\ & SPH & Energy conserving with Gaussian kernel.  & 100 & --
& --\\

\GSPH\    &GSPH & Based on Eqs. (\ref{eq:eom_gph}) and (\ref{eq:eneq_gph}) with
cubic spline for volume interpolation, & 100 & 100 \& 300 & 200 \\
       &     &second-order reconstruction and
limiter by \protect\cite{vanleer79}. & \\

\GSPHI\ &GSPH & The same as \GSPH\ but based on the limiter by
\protect\cite{inutsuka02}. & 100 & 100 \& 300 & 200 \\

\GSPHord\ &GSPH & The same as \GSPHI\ but using the first--order
reconstruction & 100 & 300 & 200 \\
          &     & for the RP solution. & & & \\

\GSPHVLin\ & GSPH & The same as \GSPH, but using the linear
interpolation for the volume  & & 300 & 200 \\
          &     & function $V(s)$, as in Eq. (\ref{eq:linvol}). & & & \\

\GSPHCW\ & GSPH & Same as \GSPH\, but based on Eqs. (\ref{eq:eom_gph}) and
(\ref{eq:eneq_gph}) by \protect\cite{cha_whitworth03}. & 100 & -- & 200\\
\hline
\end{tabular}
\end{center}
\caption{Characteristics of the hydrodynamical schemes implemented and
of the hydrodynamic tests carried out. Column 1: name of each scheme;
Column 2: hydrodynamic method on which each scheme is based; Column 3:
basic description of the implementation details of each scheme (see
Section \protect\ref{s:impl} for further details); Columns 4, 5 and 6:
number of neighbour used for each scheme in the Sod shock tube,
Kelvin--Helmholtz and blob test, respectively.}
\end{table*}

\section{Results}
\label{sec:res}
We describe in this section the tests of our GSPH implementation in
the GADGET code. The hydrodynamic tests performed are the shock tube,
the development of Kelvin--Helmholtz instabilities in a shear flow and
the disruption of a cold dense blob moving in a hot wind.
 
\subsection{Shock tube}
\label{sec:sod}
We first consider the standard Sod shock-tube test \citep{sod78}, which
provides a mean to validate the code capability to describe basic
hydrodynamic features. Initial conditions are the same used by
\cite{springel05} to test the \GADGET\ SPH scheme. An ideal gas with
politropic index $\gamma=1.4$ is considered initially at rest, filling
half space with gas at unit pressure and density ($\rho_1=1$,
$P_1=1$), and the other half space with lower density ($\rho_2=0.25$)
and lower pressure ($P_2=0.1795$) gas. Despite the intrinsic
one-dimensional nature of the test, initial conditions are generated
in three dimensions with an irregular glass-like distribution of
equal-mass gas particles. A periodic box was chosen having a longer
size in the $x$-direction, with $(L_x,L_y,L_z)=(60,1,1)$. A total
number of 75000 particles have been included in the initial conditions.
We run this test with four different hydrodynamic schemes (see Table
\ref{t:cod}), namely \GADGET, \TradSPH, \GSPH, \GSPHCW\ and
\GSPHI. In all cases, the test has been run by using 100 neighbours
within the kernel. 

We show our results for the SOD test by restricting the range of $x$
coordinates to vary between 20 and 40. In fact, since we use periodic
initial conditions, we have two shocks inside the computational
domain. We only focus on one of them, while discarding the
uninteresting unperturbed regions [0:20] and [40:50] along the
$x$-axis. Results for this test are shows as scatter plots of density,
pressure, velocity and entropy, all expressed in internal code
units. In fact, performing a binning often smooths out details which
are useful to understand the differences in the behaviours of our the
different hydrodynamic schemes.

\begin{figure*}
\includegraphics{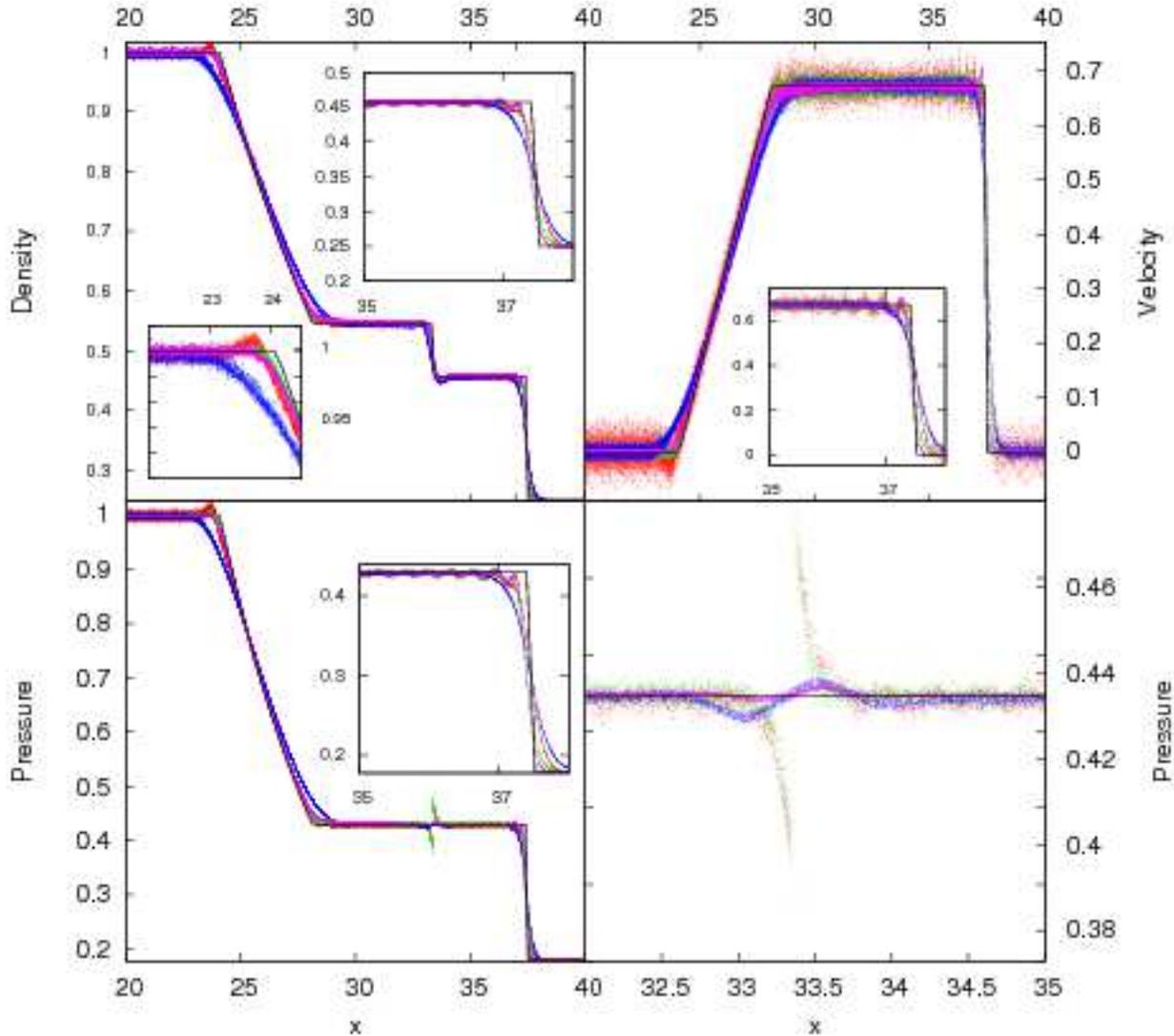}
\vspace{-3.7truecm}
\caption{Comparison between the results of different hydrodynamic
  schemes for the Sod shock tube. Scatter plots of the particle
  density, velocity and pressure are shown in the upper left, upper
  right and bottom left panels, respectively. The bottom right panel
  shows a zoom of the pressure around the contact discontinuity. Red
  points are for the entropy--conserving \protect\GADGET\ formulation,
  green points for the \protect\TradSPH\ formulation, blue points for
  the \protect\GSPHCW\ formulation and magenta points for our reference
  \protect\GSPH\ formulation (see Table \protect\ref{t:cod}). In each
  panel, the black continuous line indicates the exact solution. The
  insets emphasise the different behaviour of the four schemes, as
  also discussed in the text.}
\label{fi:sod}
\end{figure*}

We show in Figure \ref{fi:sod} density, pressure, and velocity at the
time $T=5$\footnote{Note that time is a dimensionless quantity in this
  test, thus it does not depend on our chosen systems of units}, when
the shock is well developed, for the \GADGET, \TradSPH, \GSPH\ and \GSPHCW\
schemes (red, green, blue and magenta points, respectively).

A well known problem of SPH codes in solving the Sod shock tube is the
appearance of a pressure discontinuity (``pressure blip'') at the
position of the contact discontinuity. This is at variance with
respect to the exact analytic solution, that predict instead pressure
to be continuous across this discontinuity. This pressure
discontinuity arises as a consequence of the error that the SPH scheme
makes in correctly estimating the density gradient at the density
discontinuity. As a result, a sort of spurious surface tension force
arises, due to the opposite signs that the pressure gradients have on
the two sides of the discontinuity. In fact, \GADGET\ and \TradSPH\
formulations clearly show a pressure blip at $x \approx 33.5$, where
the discontinuity is located. This is emphasised in the bottom left
panel of Fig. \ref{fi:sod}, which provides a zoom of the pressure 
in this region. Since this spurious pressure feature arises from errors
in the computation of density at the discontinuity, its origin can be
traced back to the incorrect estimate of the $1/\rho$ SPH volume
associate to each particle. Two conceptually different solutions to
this problem can be then devised. A first one is based on improving
the density estimate in the computation of density gradients across
the discontinuity \citep[e.g.][]{read10}. A second one relies instead
on the introduction of thermal diffusion across the discontinuity,
which masquerade the effect of the error in the volume estimate and
prevents the onset of the surface tension there
\citep[e.g.][]{price08}.

\cite{cha_whitworth03} noticed that the \GSPHCW\ scheme was in fact
able to prevent the development of the density blip. Since this
scheme does not pay attention to correctly estimate volumes, its
behaviour should be ascribed to the inclusion of a diffusion
term. This term is indeed provided by the $\left[\bv^*-\bx_i^*\right]$
appearing on the {\em r.h.s.} of Eq.(\ref{eq:eom_gph}), which in fact
described a net exchange of thermal energy, with a zero total mass
exchange, between each particle pair. Its effect is similar to that of
adding an artificial thermal diffusion term in the energy
equation. The main difference, however, is that here such a mixing
term naturally arises from the Godunov scheme.

As shown in the bottom--right panel of Fig. \ref{fi:sod}, we do
confirm that this scheme does not produce a pressure discontinuity,
which is replaced by an oscillation around the exact
solution. Therefore, while diffusion prevents the "pressure blip",
inaccuracy in the density estimate is still present and induces the
appearance of a spurious, although much reduced, pressure force at the
contact discontinuity. Quite interestingly, this spurious force is
further, and greatly,
reduced in our reference \GSPH\ scheme. In this case, second--order
accuracy in density estimate is enforced through the computation of
the convolution integrals in Eqs. (\ref{eq:eom}) and (\ref{eq:eneq}) (see
also I02). The net effect is a significant improvement in the
behaviour of pressure across the discontinuity.

\begin{figure}
\vbox{
\hbox{
\includegraphics[scale=0.8]{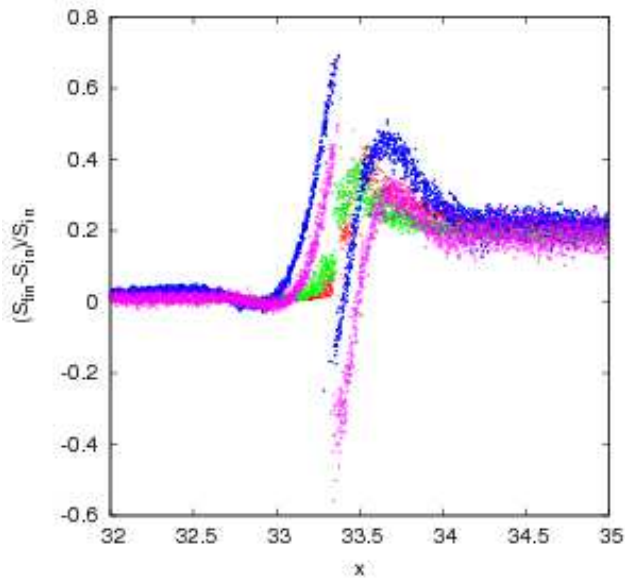}
}
}
\caption{
  Entropy variation in \protect\GADGET, \protect\TradSPH,
  \protect\GSPH\ and \protect\GSPHCW\ hydrodynamic
  scheme for the Sod shock test around the contact
  discontinuity. Colour coding
  is the same as in Figure \protect\ref{fi:sod}. Here we show the
  difference between the final and initial entropy of particles, normalised
  to their initial entropy, as a function of final particle positions. 
}
\label{fi:sod_dentr}
\end{figure}

To further emphasise the different behaviour of the SPH-based (\GADGET\
and \TradSPH) and GSPH-based (\GSPH\ and \GSPHCW) schemes, we plot in
Figure \ref{fi:sod_dentr} the fractional variation of the entropy of
particles as a function of their final position, across the
discontinuity. Quite clearly, entropy variation is only determined by
the presence of weak shocks for the \GADGET\ and \TradSPH\ scheme and,
therefore, can only have positive sign. On the other hand, the
presence of diffusion in the \GSPH\ and \GSPHCW\ allows an exchange of
internal energy across the discontinuity. This diffusion manifests
itself with both positive and negative variations of entropy, with
lower-density particles located on the right side of the discontinuity
loosing thermal energy in favour of particles located on the other
side of the discontinuity. It is such a mixing that is responsible of
the damping of the pressure blip.

Both \GADGET\ and \TradSPH\ schemes show pressure wiggles immediately
before the shock wave, which is located at $35<x<37$ (see the inset in
the pressure panel in Fig. \ref{fi:sod}).  Such wiggles in pressure
correspond to wiggles in the velocity, as shown in the top-right panel
of Fig.\ref{fi:sod} at the same position. Note that velocity wiggles
are largest for \GADGET\ and smallest for \GSPH. Wiggles are also
present in the \GADGET\ run at the same position for the density
variable (upper-left panel). The insets in the two panels show a
blow-up of the region around the shock wave positions, and emphasise
the presence of such wiggles. The prominence of the pressure wiggles
in the \GADGET\ simulation is due to the lack of thermal energy
diffusion that characterises this hydrodynamic scheme. In a sense,
these wiggles have the same origin as the wiggles in velocity
appearing in the shock-tube test when artificial viscosity is not
included \citep[e.g.,][]{rosswog09}. In fact, artificial viscosity
removes velocity wiggles since it effectively provides a diffusion of
momentum at the shocks. In a similar way, diffusion of thermal energy
associated to the GSPH scheme is effective in removing wiggles in
pressure at such discontinuities.

In the upper-right panel of Figure \ref{fi:sod}, we also notice that
the scatter in the velocities is minimum for \GSPH\ and maximum for
\GADGET. The shock wave is better captured by \GADGET\ and \GSPH, with
\GSPHCW\ performing worse. It is quite remarkable that \GSPH\ is able
to correctly capture the shock, while preventing at the same time the
appearance of noise in the velocity field, without the introduction of
an artificial viscosity term in the momentum equation. Furthermore,
the better accuracy of \GSPH\ with respect to \GSPHCW\ is the
consequence of the improved accuracy of the former scheme.

The shape of the rarefaction fan in density and pressure (upper and
lower left panels in Fig. 1, respectively) are better captured by
\GADGET\ and \TradSPH\ formulations, while \GSPHCW\ shows a much smoother
behaviour. \GSPH\ also shows a slight smoothing of density and pressure
at the beginning and at the end of the rarefaction fan. On the other
hand, \GADGET\ produces a piling-up of particles at the onset of the
fan, corresponding to an excess of particles having low velocities,
at the same position. This feature can be better appreciated in the
lower inset of upper-left panel of Fig. 1.

\begin{figure*}
\vbox{
\hbox{
\includegraphics{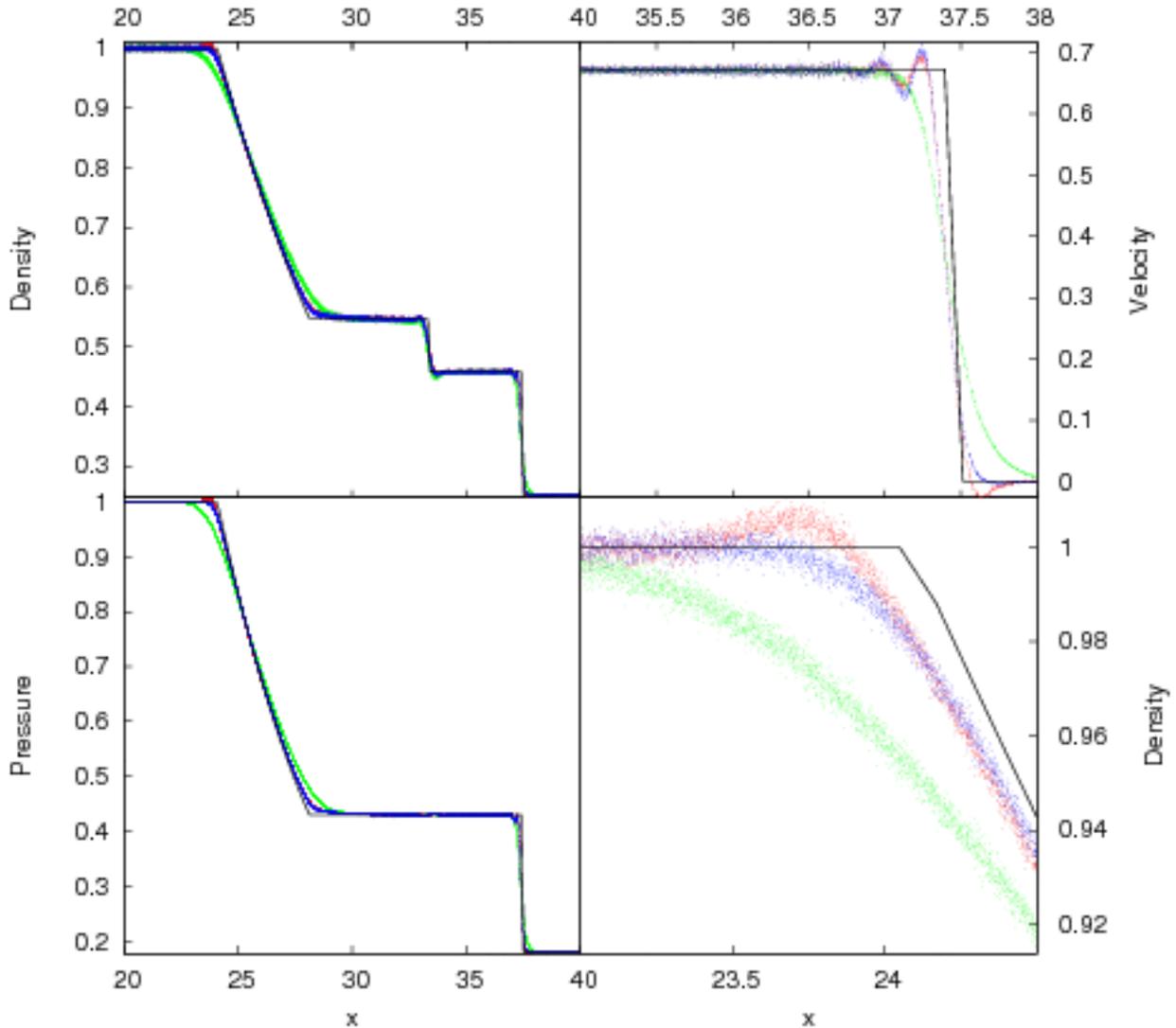}
}
}
\vspace{-3.7truecm}
\caption{Comparison between the results of different implementations
  of the GSPH Equations (\protect\ref{eq:eom}) and
  (\protect\ref{eq:eneq}) for the Sod shock tube. Scatter plots of
  particle density, velocity and pressure are shown in the upper left,
  upper right and bottom panels, respectively. The bottom-right panel
  shows a zoom-in of pressure behaviour around the onset of the
  rarefaction fan. Red points are for the \protect\GSPH\ with the
  standard limiter of \protect\cite{inutsuka02} (\protect\GSPHI),
  green points for the same scheme when using a first-order
  reconstruction of density and pressure at the interface in the
  solution of the Riemann problem (\protect\GSPHord), while the blue
  points correspond to our standard implementation (\protect\GSPH)
  based on using the limiter by \protect\cite{vanleer97} and the
  third-order spline interpolation of volumes.  }
\label{fi:sod_gsph} 
\end{figure*}

Figure \ref{fi:sod_gsph} shows results for the Sod test, when we use
different implementation for the \GSPH\ formulation. Besides our
standard implementation of \GSPH, we also show results based on the
limiter used by I02 in its implementation of GSPH (\GSPHI) and on the
first--order reconstruction at the interface $s_*$ for the solution of
the RP (\GSPHord; see Table\ref{t:cod}). From the behaviour of density
and pressure, it is clear that a first-order reconstruction results in
a more diffusive behaviour: the shape of the rarefaction ramp and of
the velocity profile at the position of the shock wave are smoother
than for the other two implementations. For the same reason, no
velocity wiggles appears when this formulation is used. The
difference between the two limiters is instead clear when we analyse
the density right before the rarefaction fan. There, the standard
limiter produces an accumulation of particles, similar to what we
found when using the SPH \GADGET\ formulation. Such an accumulation is
instead eliminated by adopting the limiter by
\cite{vanleer79}. Furthermore, the use of this limiter avoids the
accumulation of low-velocity particles after the position of the shock
wave, as can be seen in the upper-right panel. The pressure blip at
the density discontinuity is erased in all of these three schemes,
thus we do not show a zoom-in for the pressure.

\subsection{Kelvin-Helmholtz instability} 
\label{sec:KH} 
The Kelvin-Helmholtz instability occurs across a contact
discontinuity in the presence of a tangential shear flow. It belongs
to a class of tests that have been used over the last few years to
assess the capability of SPH and grid-based methods to capture
fluidodynamical instabilities
\citep[e.g.][]{agertz07,price08,wadsley08,read10,valcke10}. It
describes the evolution of two fluids having different densities and
in pressure equilibrium, moving with opposing velocities. The
interface between the fluids is perturbed leading to a phase in which
fluid layers develop vortex instabilities, with subsequent mixing of
the two phases \citep{chandra61}. We remind here that a
two--dimensional version of this test has been also recently discussed
in detail by \cite{cha10} to assess the capability of GSPH to describe
instabilities. We extend here this test in three-dimensions for our
GADGET implementation of GSPH.

The KH test that we present here belongs to the ``Wengen'' suite of
hydrodynamical tests\footnote{\tt http://www.astrosim.net/code}, and
is described in detail by \cite{read10}. An ideal idea gas with
politropic index $\gamma =1.4$ and mean molecular weight $\mu=1$ is
assumed. Initial conditions are generated in the periodic simulation
domain $(L_x,L_y,L_z)=(256,256,16)$ kpc centered on the origin. Two
domains with $|y|<64$ kpc and $|y|>64$ kpc corresponds to the two
fluids having $\rho_1=2\rho_2$, $T_1=0.5T_2$ and opposing velocities
with the same modulus $v$. In this test, $\rho_2=3.13 \cdot
  10^{-8}$ M$_\odot$ kpc$^{-3}$, $T_2=3\cdot 10{^6}$K, and $v=40$ km
  s$^{-1}$.  Equal-mass gas particles are initially located on a
grid, whose spacing in the two domains is set in such a way to
reproduce the difference in density. Instabilities are triggered by
imposing a velocity perturbation along the $y$-direction, having a
characteristic wavelength $\lambda = 128$ kpc. The characteristic KH
time-scale for the development of instabilities from this perturbation
is
\be \tau_{KH}={\lambda (\rho_1+\rho_2)\over 2v(\rho_1
  \rho_2)^{1/2}}\,.  \label{eq:tau_kh} 
\ee 

The units used in the Wengen tests are: kpc for length, km s$^{-1}$
for velocities, and $10^{10}$ M$_\odot$ for masses. In this system,
the unit of time is $t_*=0.977$ Gyr. Using the perturbation described
above, $\tau_{KH} \simeq 3.32$ Gyr. Initial conditions have been
generated using 774.144 gas particles. We carried out this test using
different implementations of the GSPH scheme: our reference scheme
(\GSPH), the scheme based on the limiter adopted by I02 (\GSPHI), the
scheme based on a first-order reconstruction for the thermodynamical
quantities at the interface (\GSPHord), and the scheme based on the
linear interpolation of the volume function $V(s)$ (\GSPHVLin).
In order to reproduce the results reported for the \GADGET\ code on
the web site of the Wengen tests, we carried out the KH test in this
case using 442 neighbours. As for the different implementations of the
GSPH scheme, we always used 300 neighbours within the Gaussian kernel,
while we also checked the effect of using instead 100 neighbours for
the reference \GSPH\ scheme and for the \GSPHI\ scheme.

\begin{figure*}
\hspace{0.8truecm}
\vbox{
\hbox{
\includegraphics[scale=0.25,angle=90]{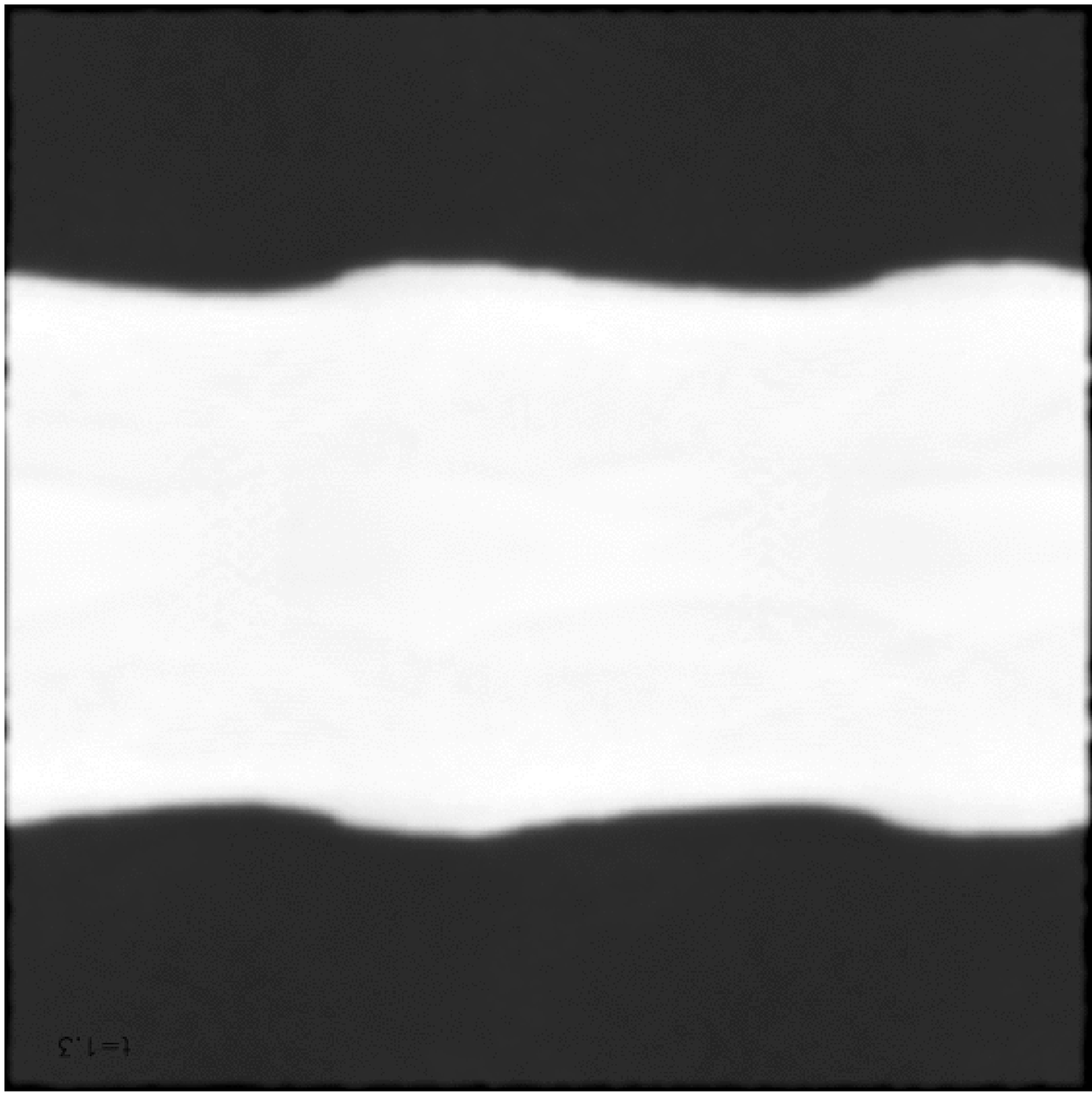}
\includegraphics[scale=0.25,angle=90]{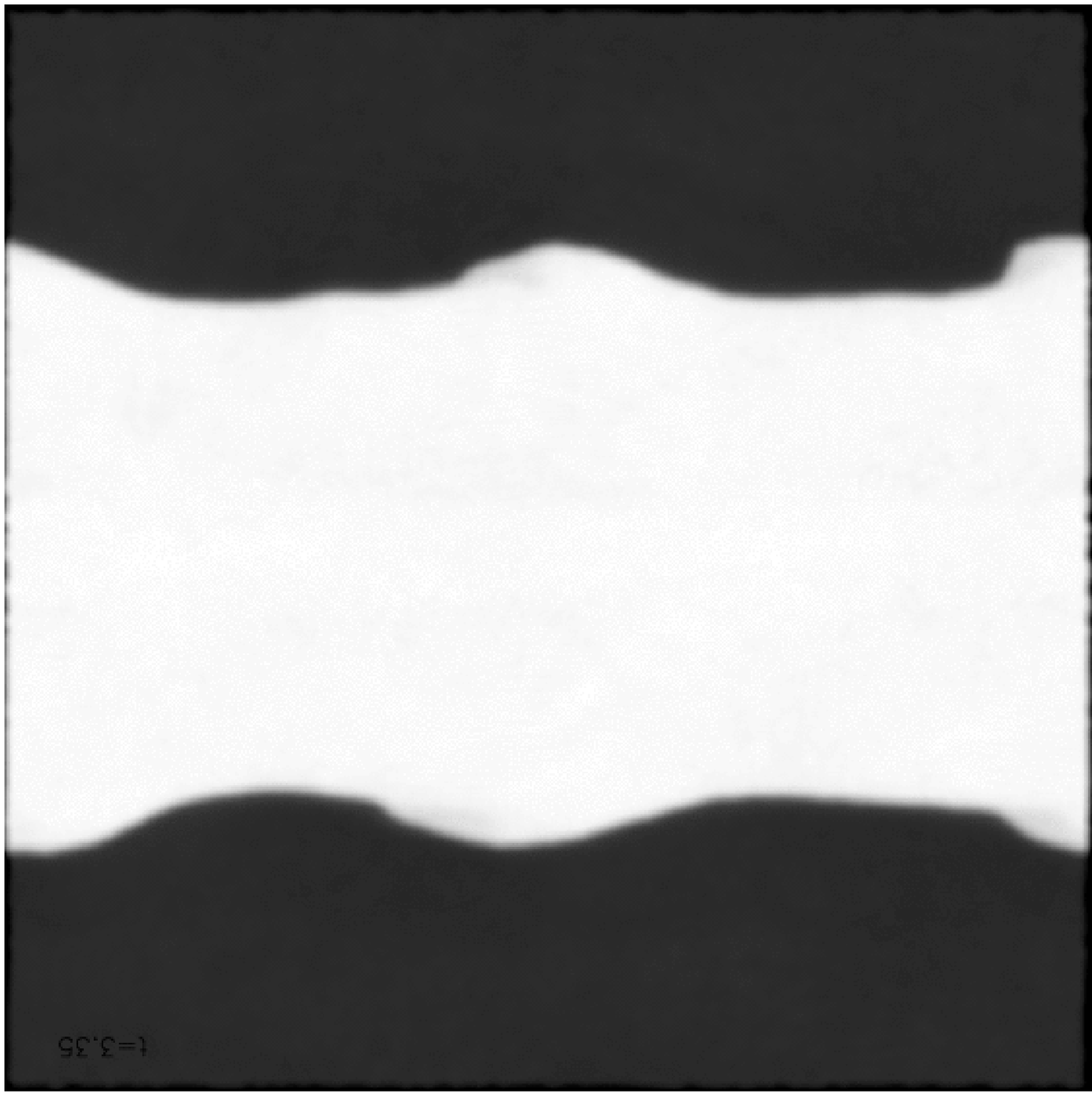}
\includegraphics[scale=0.25,angle=90]{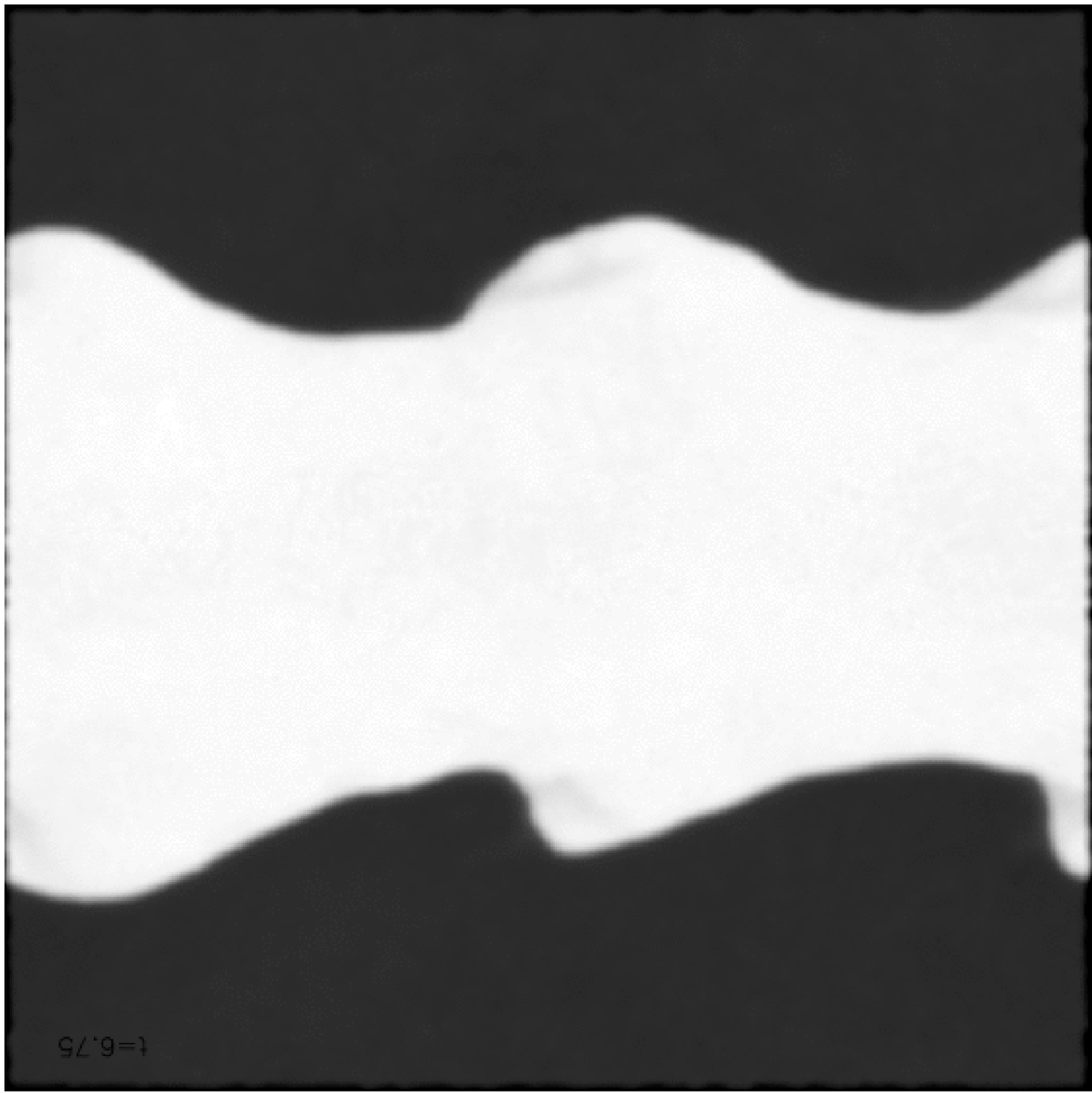}
}
\hbox{
\includegraphics[scale=0.25,angle=90]{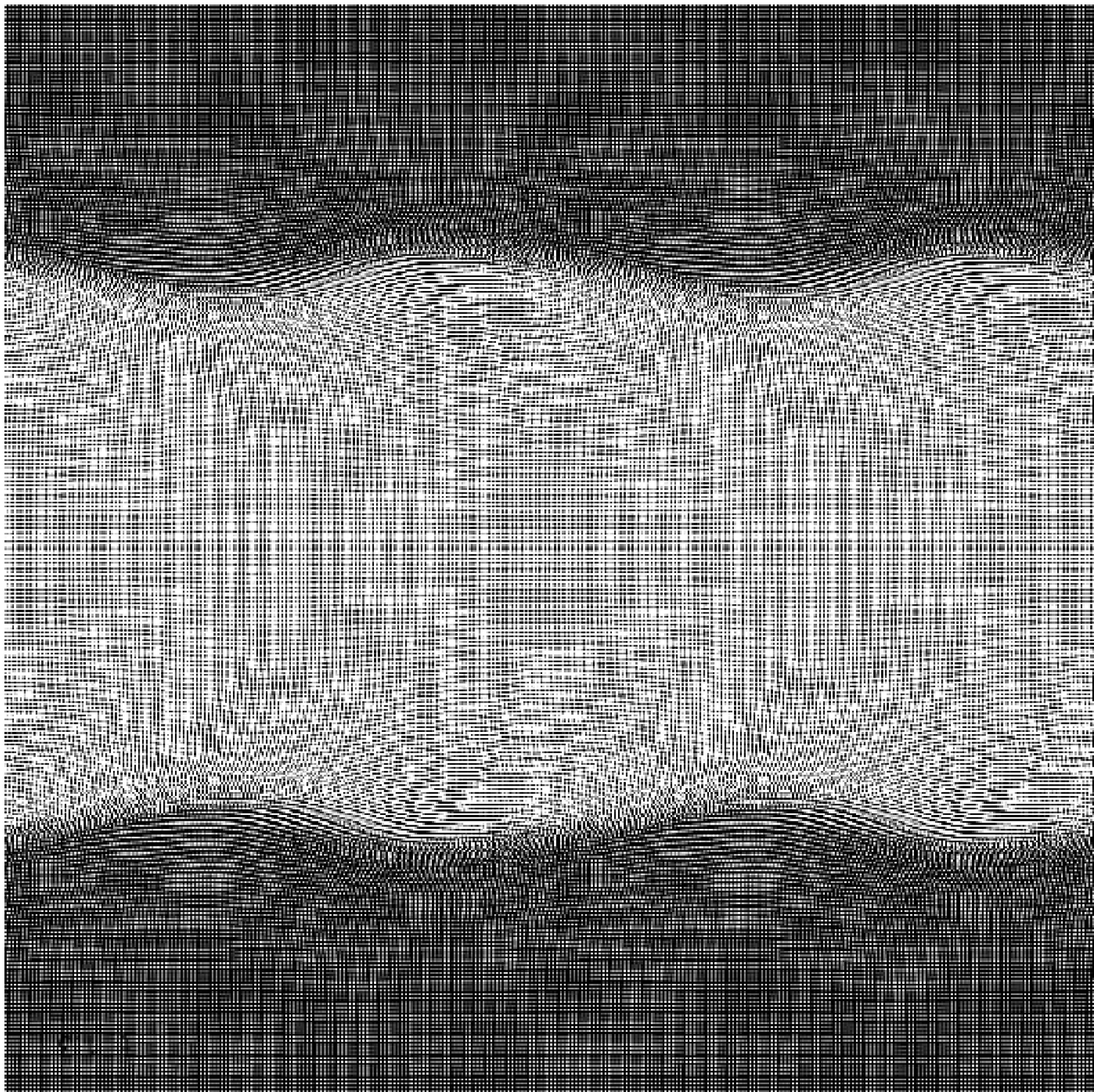}
\includegraphics[scale=0.25,angle=90]{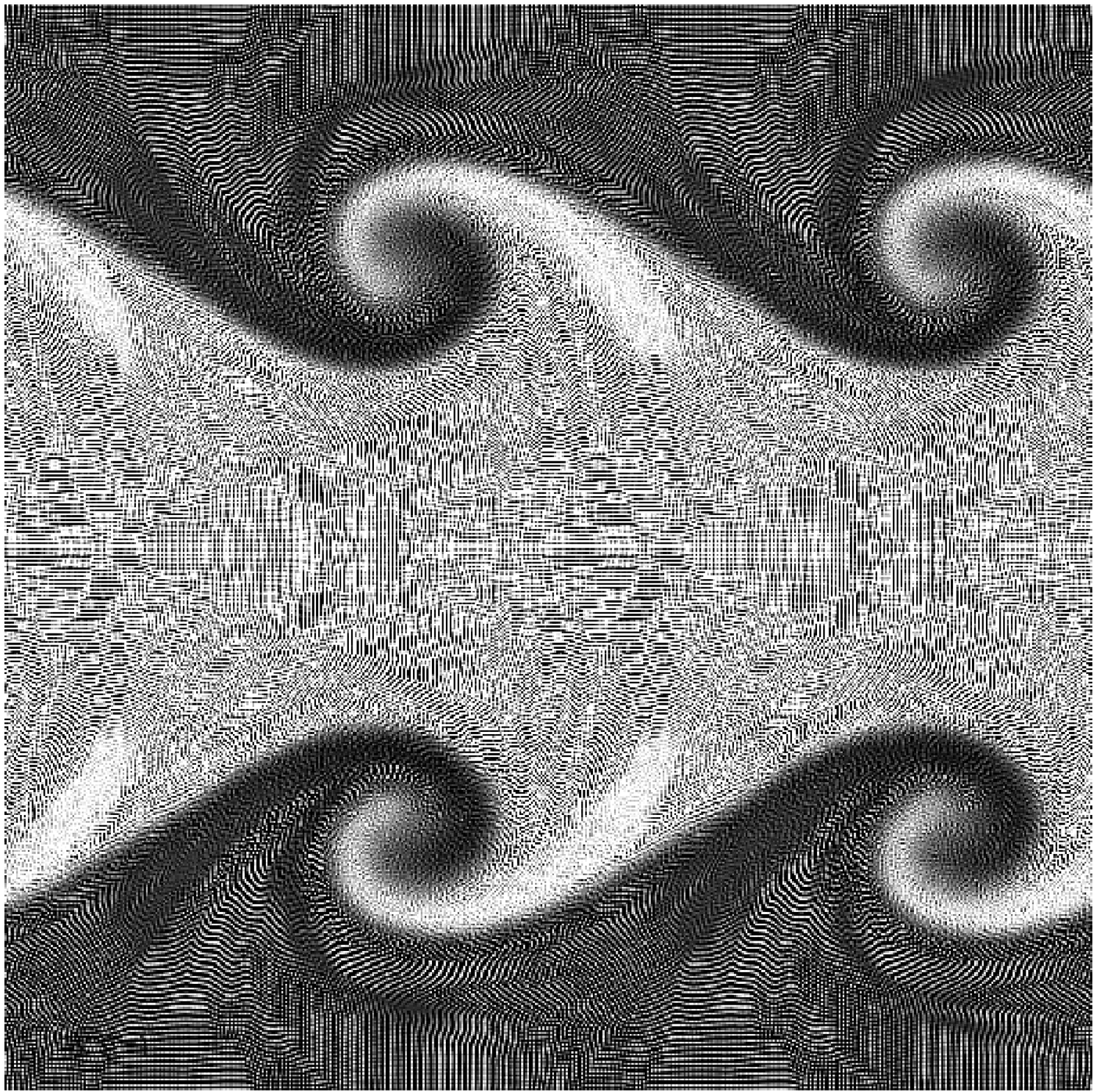}
\includegraphics[scale=0.25,angle=90]{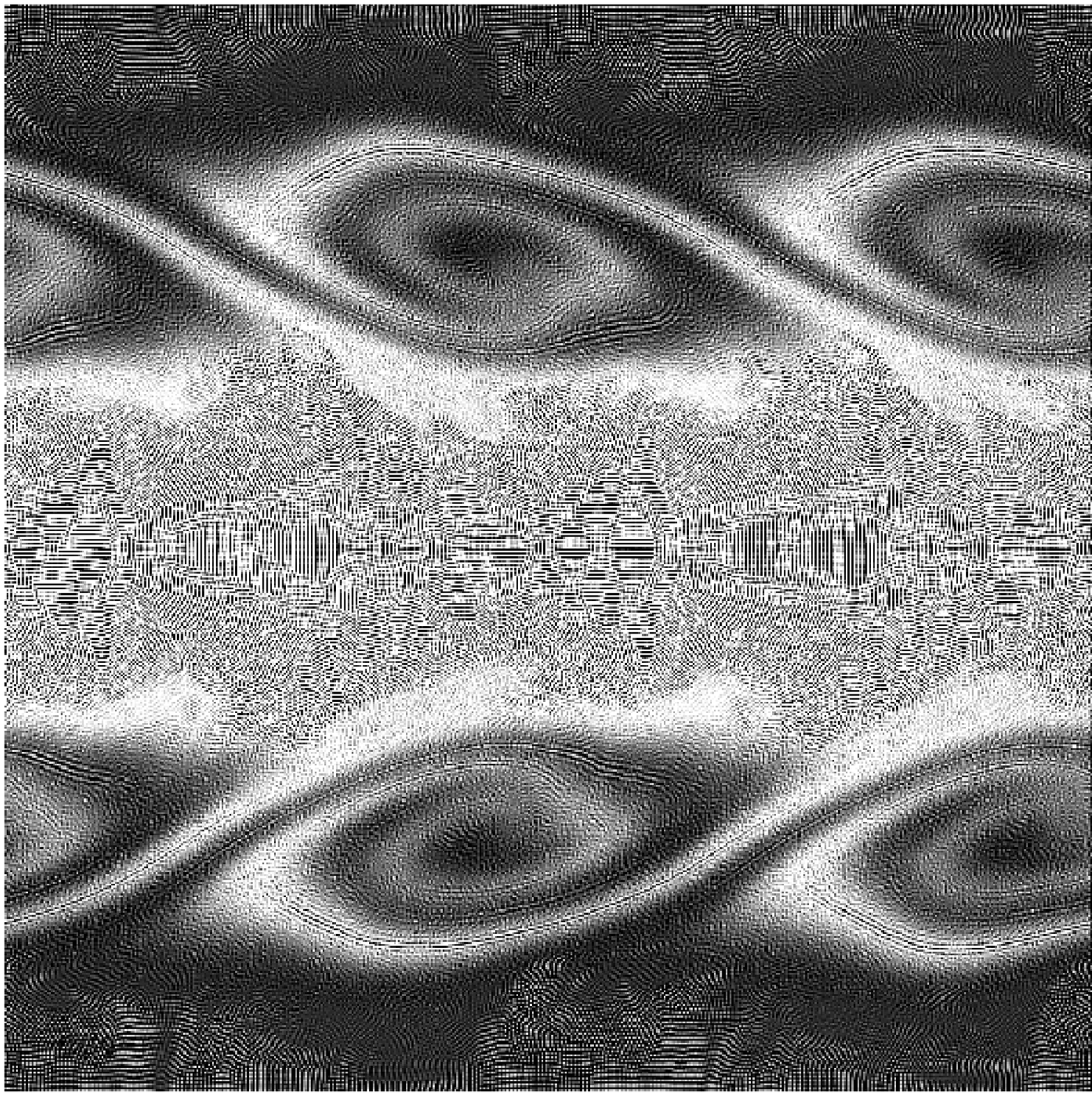}
}
\hbox{
\includegraphics[scale=0.25,angle=90]{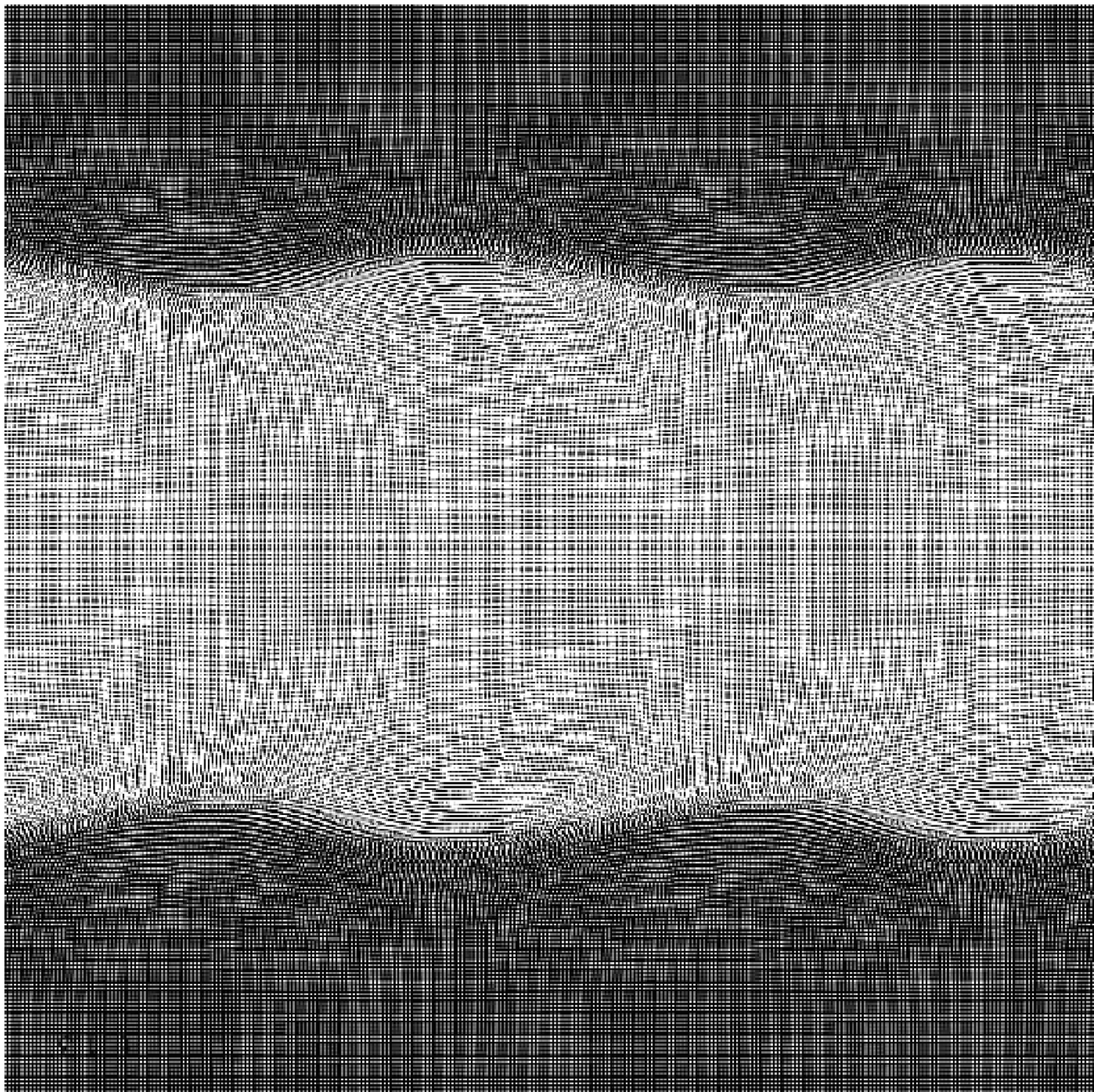}
\includegraphics[scale=0.25,angle=90]{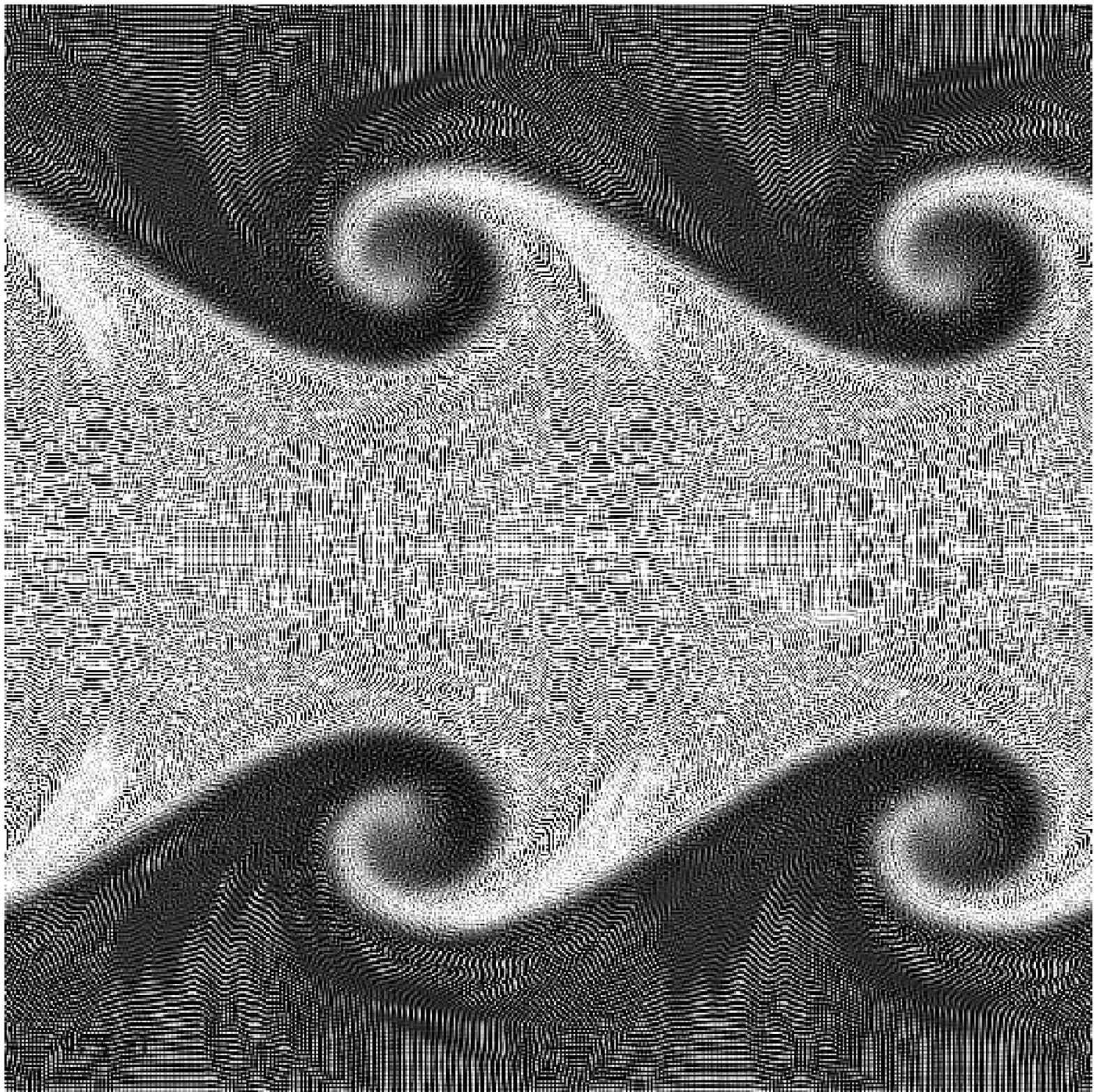}
\includegraphics[scale=0.25,angle=90]{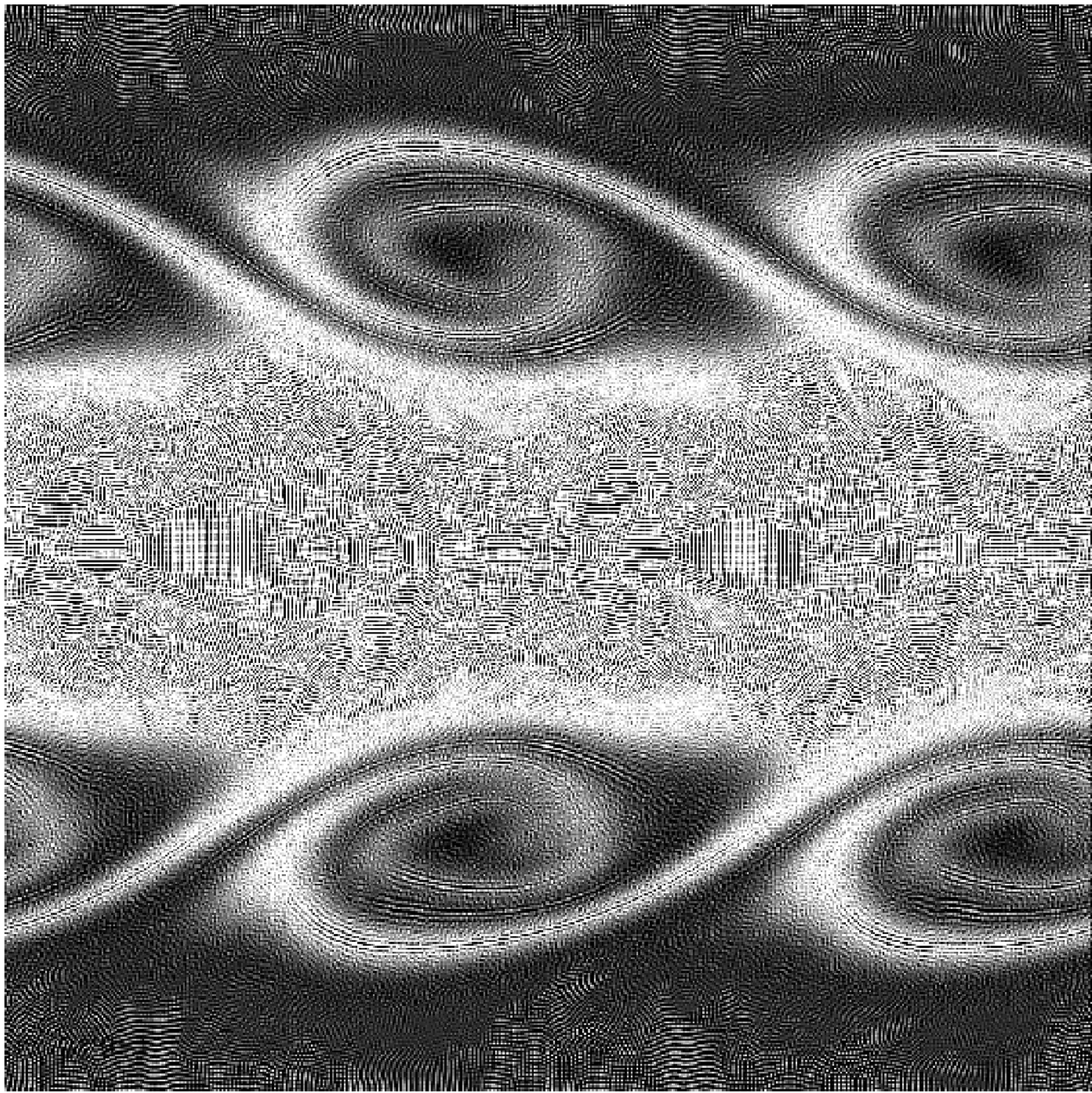}
}
}
\caption{Development of the Kelvin-Helmholtz instabilities at different
  times, as described by the \protect\GADGET\ (upper panels),
  \protect\GSPHI\ (central panels) and \protect\GSPH\ (bottom panels)
  scheme. Results are shown at three different times,
  $0.5\tau_{KH}$,$\tau_{KH}$ and $2\tau_{KH}$ in the left, middle and
  right panels, respectively. Each panel report the the projected
  density for a slice of coordinates $-3.5<z<5.5$ kpc.  Colour scale
  is linear and ranges from $2 \times 10^{-7}M_\odot$kpc$^{-2}$
  (black) to $5\times 10^{-7}M_\odot$kpc$^{-2}$ (white).}
\label{fi:kh}
\end{figure*}

We show in Figure \ref{fi:kh} the development of the KH instability
for \GADGET, \GSPH\ and \GSPHI at three different times, namely
$0.5\tau_{KH}$,$\tau_{KH}$ and $2\tau_{KH}$. It is clear that, while
the standard entropy-conserving SPH scheme dumps the instability, both
our \GSPH\ and \GSPHI\ schemes successfully capture its development. At
$t=\tau_{KH}$, vortexes begin to show up, and at $t=2\tau_KH$ they are
fully developed and display the typical ``cat-eye'' structures in gas
density.

This figure confirms the results reported by \cite{cha10} and
demonstrates the capability of the GSPH scheme to develop
gas--dynamical instabilities. Note that, while the damping of the
instability in the \GADGET\ entropy conserving scheme is due to the
use of an artificial viscosity and to development of a ``artificial
surface tension'' force at the interface, originated by the repulsion
of SPH particle at the discontinuity (see e.g \cite{price08,cha10}),
the former is absent and the latter strongly reduced in GSPH schemes
(see \cite{inutsuka02,cha10} for a discussion of the different
estimate of density on GSPH and on its effect on the artificial
surface tension).  The much improved description that the Godunov SPH
scheme provides in describing the development of KH instabilities has
to be ascribed to the fact that this method is based on explicitly
computing the convolution integrals in Eqs. (\ref{eq:eom}) and
(\ref{eq:eneq}), to a $O(h^2)$ accuracy, through the interpolation of
the volume function $V(s)$.

\begin{figure*}
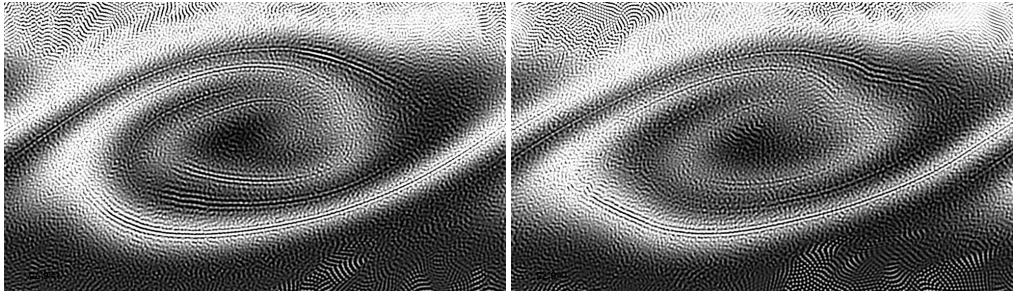

\hspace{1.truecm}
\hbox{
\includegraphics[scale=0.25,angle=90]{GSPH-Vllim-300nn-T6.75.b.ZOOM.ps}
\includegraphics[scale=0.25,angle=90]{GSPH-300nn-T6.75.b.ZOOM.ps}
}
\caption{Blow-up of the curl structure developed by the KH
  instability at $t=6.75$, for \protect\GSPH\ (left panel) and by
  \protect\GSPHI\ (right panel). The structure shown here developed in
  the ranges of coordinates $78<x<210$ kpc and $161<y<237$
  kpc.  We show the projected density for a slice of coordinates
  $-3.5<z<5.5$ kpc.  Colour scale is linear and ranges from $2 \times
  10^{-7}M_\odot$kpc$^{-2}$ (black) to $5\times 10^{-7}
  M_\odot$kpc$^{-2}$ (white).}
\label{fi:kh_zoom}
\end{figure*}

Figure \ref{fi:kh_zoom} shows a blow-up of one of the curl
  structures, which forms after the KH instability is fully
  developed, for both the \GSPH\ and the \GSPHI\ schemes. The
  difference is small, but the Van Leer limiter is able to more neatly
  capture the vortex structure. We hypotize that the reason for this
  behaviour is that the Van Leer limiter is capable to slightly reduce
  the intrinsic residual numerical diffusion associated to the Riemann
  solver, with respect to the limiter implemented by I02.

\begin{figure}
\includegraphics[scale=0.8]{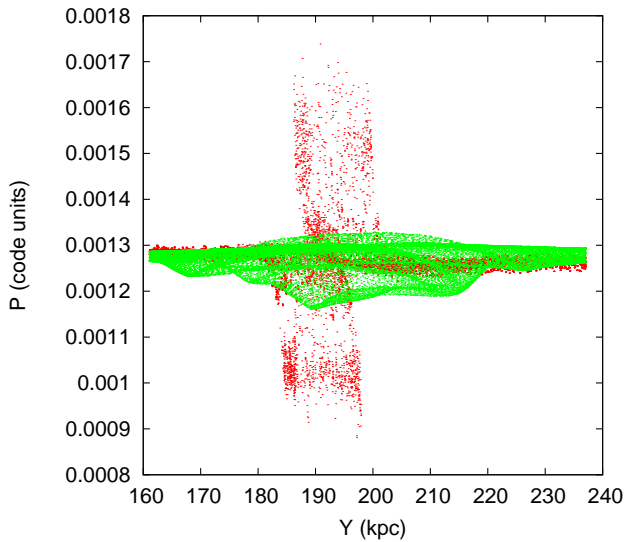}
\caption{Pressure of gas particles as a function of their $y$
  coordinates for \protect\GADGET\ (red dots) and \protect\GSPH\
  (green dots), at time $t=2\tau_{KH}$, in correspondence of the same vortex
  shown in Fig. \protect\ref{fi:kh_zoom}, $78<x<210$ kpc, $161<y<237$
  kpc, $-3.5<z<5.5$ kpc.  }
\label{fi:kh_pres}
\end{figure}

To further highlight the different ways in which entropy--conserving
SPH and \GSPH\ respond to the velocity perturbation across the contact
discontinuity in the KH test, we show in Figure \ref{fi:kh_pres} a
scatter plot of the pressure of gas particles as a function of the $y$
coordinate (i.e. in the direction parallel to the direction to the
velocity perturbation triggering the KH instability). This scatter plot
includes only the particles contained within the same vortex region
shown in Fig. \ref{fi:kh_zoom}. It is clear that \GADGET\ develops a
pressure ``blip'' at the discontinuity, similarly to what happens in
the shock tube test. On the contrary, the blip basically disappeared
in the \GSPH\ scheme, and a complex pressure structure, associated to
the vortex, develops.

\begin{figure*}
\hspace{0.truecm}
\vbox{
\hbox{
\includegraphics[scale=0.25,angle=90]{GSPH-Vllim-300nn-T6.75.b_lr.ps}
\includegraphics[scale=0.25,angle=90]{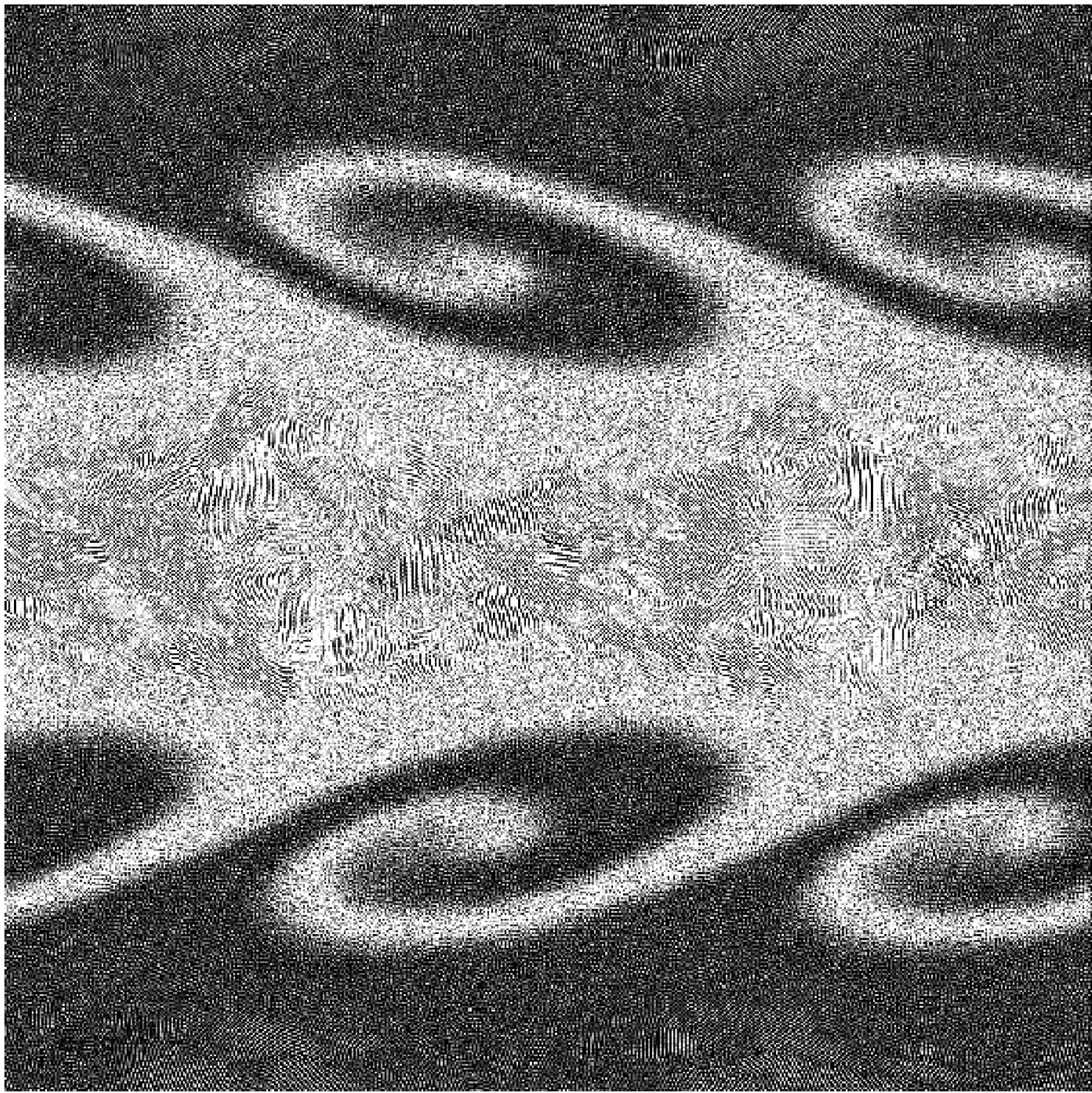}
\includegraphics[scale=0.25,angle=90]{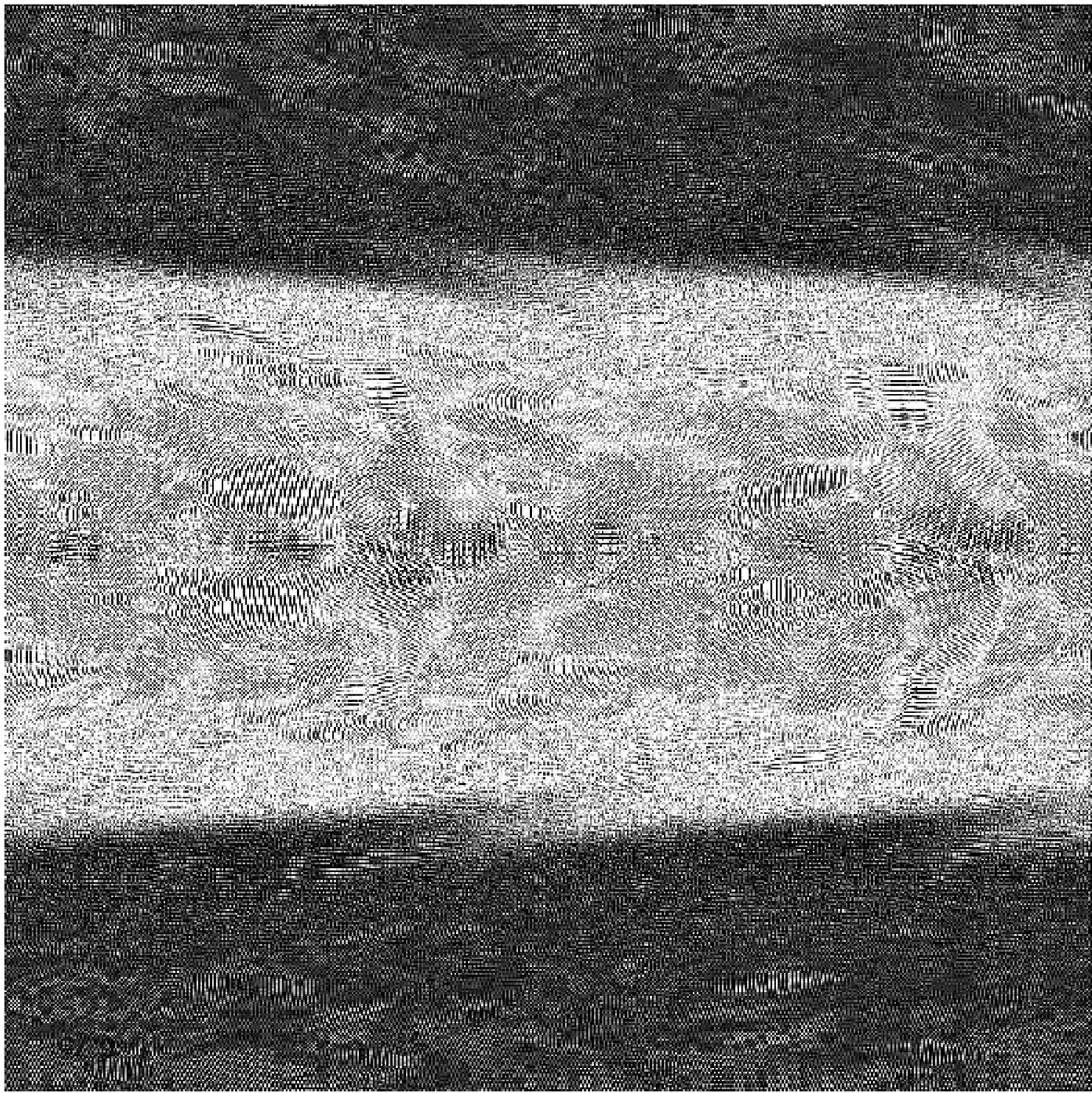}
}
\hbox{
\includegraphics[scale=0.25,angle=90]{GSPH-300nn-T6.75.b_lr.ps}
\includegraphics[scale=0.25,angle=90]{kh_vlin300nn_135.ps}
\includegraphics[scale=0.25,angle=90]{kh_gsph_cw_135.ps}
}}
\caption{Effect of different GSPH implementations on the KH
  instability at $t=2 \tau_{KH}$. Results are shown for our reference
  \protect\GSPH\ scheme using 300 neighbours (upper left panel) and
  100 neighbours (upper central panel), using \protect\GSPHord\ with
  first-order reconstruction at the interface (upper right panel), and
  using \protect\GSPHI\ with the limited by I02 (bottom left panel),
  using \protect\GSPHVLin\ with linear interpolation for the volume
  function (bottom central panel), and for the \protect\GSPHCW\ scheme
  based on Eqs. (\protect\ref{eq:eneq_gph}) and
  (\protect\ref{eq:eom_gph}) (bottom right panel). Ranges of
  coordinates and gray-scale density coding are the same as in
  Fig. \protect\ref{fi:kh}.}
\label{fi:kh_gsph}
\end{figure*}

In order to further highlight the role that different details in the
implementation of the GSPH scheme have in the development of the KH
instability, we also studied the effect of varying the number of
neighbours, the reconstruction prescription and the volume
estimates. The results of these tests are shown in Figure
\ref{fi:kh_gsph}.  The most striking effect is given by the
reconstruction order for the thermodynamical quantities at the
interface where we solve the Riemann problem. A first-order
reconstruction (\GSPHord; upper right panel) is so diffusive that the
KH instability does not develop at all. This demonstrates how crucial
it is for the development of the KH instability to choose a scheme
that gives the lowest possible degree of diffusion where it is not
required. In fact, the \GSPHord\ has been shown to be effective in
preventing the formation of the pressure blip in the shock tube test
(see Fig.\ref{fi:sod}). However, the excess of diffusion, which
manifests itself in the shock tube test as a smooth transition at the
rarefaction fan, is such to prevent the development of the KH
instability. We also note that it is quite important to use a rather
large number of neighbours: decreasing its number from 300 to 100
(upper central panel) produces a smoother structure of the vortexes,
and fails to capture their ``cat-eye'' shape. This is due to the
increase of noise in the density estimate at the discontinuity, that
arises when a smaller number of neighbours is used. On the other
hand, the KH test is also sensitive to the precision of interpolation
of the volume function: using a linear interpolation of this function
(\GSPHord; lower central panel), develops instabilities whose ``curl''
structure is however less resolved than for a cubic interpolation. As
for the \GSPHCW scheme, based on Eqs. (\ref{eq:eom_gph}) and
(\ref{eq:eneq_gph}), the large amount of diffusivity is such to
prevent the development of KH instabilities. The lack of accuracy of
this scheme is due to two main differences with respect to our
reference GSPH scheme: firstly, Eqs. (\ref{eq:eom_gph}) and
(\ref{eq:eneq_gph}) are not derived from the convolution of the
equation of motion and energy conservation; secondly, in this scheme we
have no means to locate the position of the interface for the solution
of the RP, thus implying that we are effectively resorting to a
first-order reconstruction.

\begin{figure*}
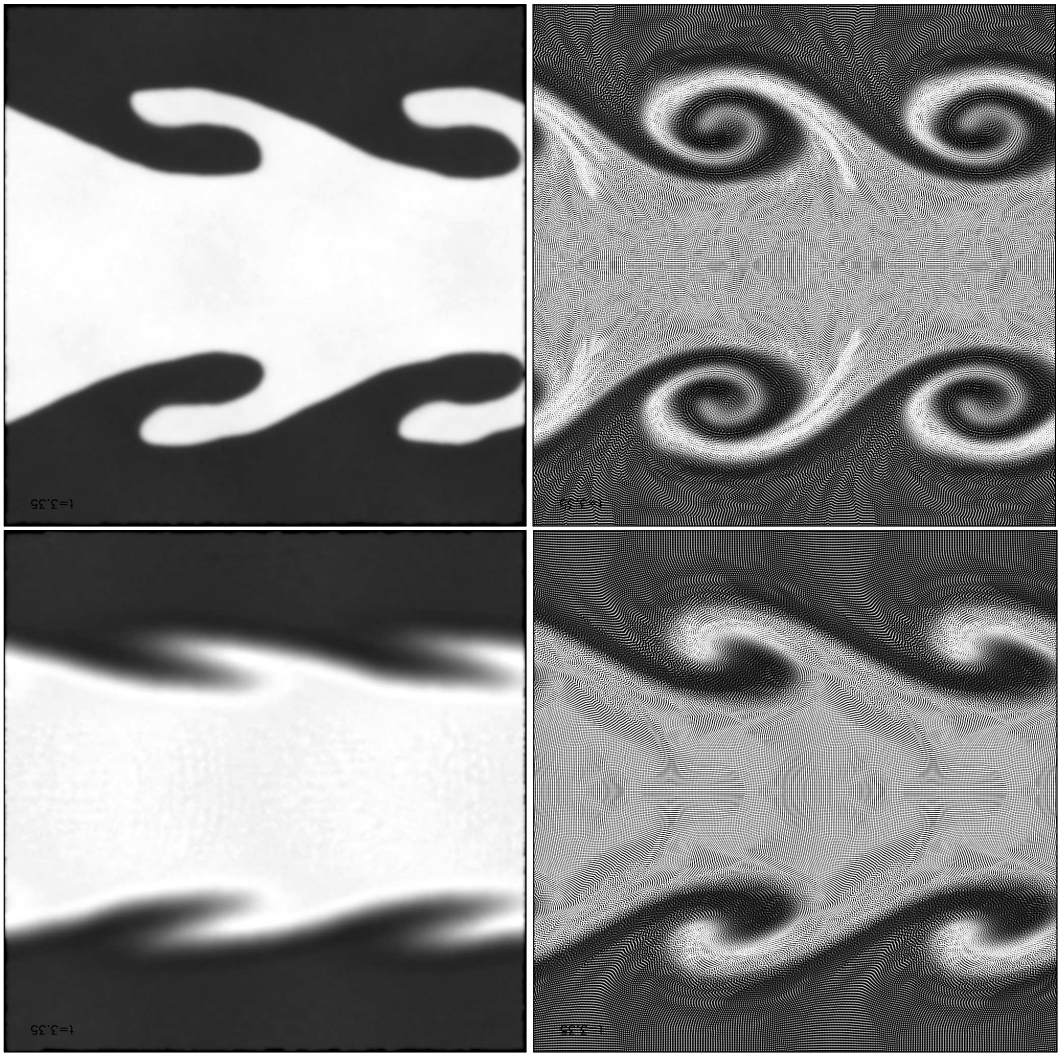

\hspace{0.truecm}
\vbox{
\hbox{
\includegraphics[scale=0.35,angle=90]{GADGET-LargeVy-067.ps}
\includegraphics[scale=0.35,angle=90]{GSPH-300nn-LargeVy-067.ps}
}
\hbox{
\includegraphics[scale=0.35,angle=90]{kh_gsph_cw_LargeVy_067.ps}
\includegraphics[scale=0.35,angle=90]{GSPH-Vcub-norec-LargeVy-067.ps}
}}
\caption{
  KH test with a stronger velocity perturbation on the y axis.
  We show the result at the time $\tau_{KH}=1$ of our standard KH test.
  Results are shown for the entropy conserving \protect\GADGET\ scheme (upper
  right panel), for our reference one  \protect\GSPH\ (upper left panel),
  for \protect\GSPHCW\ (lower left panel) and for
  \protect\GSPHord\ (lower right panel). Ranges of
  coordinates and gray-scale density coding are the same as in
  Fig. \protect\ref{fi:kh}.
}
\label{fi:kh_largevy}
\end{figure*}

As a further check, we run another KH test, in which we
  multiplied the y-axis velocity perturbation by a factor of five. The
  aim of this test is to verify whether lower-order schemes succeed in
  following the development of the KH instability when the
  perturbation is strong enough. We show the results of this test in
  Figure \ref{fi:kh_largevy}, at the time $t=\tau_{KH}$ of our
  ``standard'' test, so that also the amplification of the
  perturbation is appreciable. In this case, the entropy conserving
  \GADGET\ scheme develops arms, but fails to capture the development
  of the vortexes.  On the other hand, \GSPH\ is confirmed to
  successfully capture the instability, while the \GSPHord\ and
  \GSPHCW\ schemes confirm to be very diffusive and to smooth out the
  instability. The results of this test demonstrate that a
  higher-order \GSPH\ scheme is necessary to correctly treat the
  development of KH instabilities. 

In summary, the results shown in this Section confirm and extend the
2D test results presented by \cite{cha10} on the capability of GSPH to
follow the development of KH instabilities. We demonstrated that the
performance of GSPH is further improved by adopting the limiter by
\cite{vanleer79}, instead of that of I02, used by
\cite{cha10}. Furthermore, our results also highlight that the
development of the instability is inhibited in different ways by
{\em (a)} errors in density estimate at the discontinuity, as in
standard SPH or when using a small number of neighbours in GSPH, and
{\em (b)} numerical diffusion, which increases when using a less
accurate first-order reconstruction of thermodynamical variables at
the interface.

\subsection{The blob test}
\label{sec:blob}
This test describes the disruption of a cold gas cloud having uniform
density, moving in pressure equilibrium against a hot lower-density
wind. This test has been used by \cite{agertz07} to assess the
capability of different hydrodynamic schemes to describe the blob
disruption due to the onset of KH and Rayleigh--Taylor instabilities
\citep[see also][]{read10}. \cite{cha10} recently used a
two--dimensional version of this test to assess the performance of
their GSPH implementation. The version of the blob test that we use
here also belongs to the ``Wengen'' suite. A full description of the
initial conditions is provided by \cite{read10}. The simulation domain
is given by a periodic rectangular box with $(L_x,L_y,L_z)=(2,2,8)$
Mpc, with the origin of the coordinates located at the centre of this
domain and the blob initially located at $(0,0,-3)$ Mpc. The radius of
the cloud is set to $r_c=197$ kpc. Internal density of the cloud is a
factor $\chi=10$ higher that in the external medium, with the
temperature being correspondingly a factor 10 lower so as to fulfil
the condition of pressure equilibrium. Initial conditions are
generated by placing equal-mass particles in a lattice configuration,
so as to satisfy the above density requirements. The velocity of the
wind is $v=1000\,\vel$. An initial instability is also used to the
surface layer of the cloud to trigger a large-scale instability
\citep[see][ for a full description of the initial
conditions]{read10}. Units of measure for length, velocity, mass and
time are the same as in the KH test. Initial conditions are generated
using $10^7$ gas particles. We use 200 neighbours for the test based
on the different implementations of GSPH, and 50 for GADGET, which
uses the B--spline kernel (see Table \ref{t:cod}). Following
\cite{agertz07}, we define the cloud crushing time, $\tau_{cc}={r_c
  \sqrt \chi \over v} = 0.61$ Gyr, which gives the typical time--scale
for the evolution of the cloud moving at supersonic velocity.

The results of this test have implications for a number of relevant
astrophysical and cosmological applications, in which a dense gas
cloud interact with a lower density medium. For instance, this is the
case of a cold molecular cloud in the inter-stellar medium, which
interacts with ejecta from a nearby exploding supernova. Another
example is provided in cosmological simulations by substructures
bringing relatively cold gas which merge into larger halos permeated
by hotter gas during the hierarchical assembly of galaxy groups and
clusters.

The blob is expected to be initially destabilised by
Richtmyer-Meshkov instability (RM) and by
Rayleigh--Taylor (RT) instability and subsequently further dissolved
by KH instability. 
We show in Figure \ref{fi:blob} the evolution of
the projected gas density in this test at three different times, for
the GADGET (upper panels), \GSPH\ (bottom panels) and \GSPHI\ (middle
panels) hydrodynamical schemes. As expected, the limitations of the
standard entropy conserving formulation of SPH in following
hydrodynamical instabilities make this scheme unable to describe the
disruption of the cold blob \citep[see][for a detailed
discussion]{agertz07}. Already at early times, $t=4$ (left panels),
the \GADGET\ simulation develops less 
hydrodynamical instabilities in the up-wind part of the blob.

\begin{figure*}
\hspace{1truecm}
\vbox{
\hbox{
\includegraphics[angle=-90,scale=0.45]{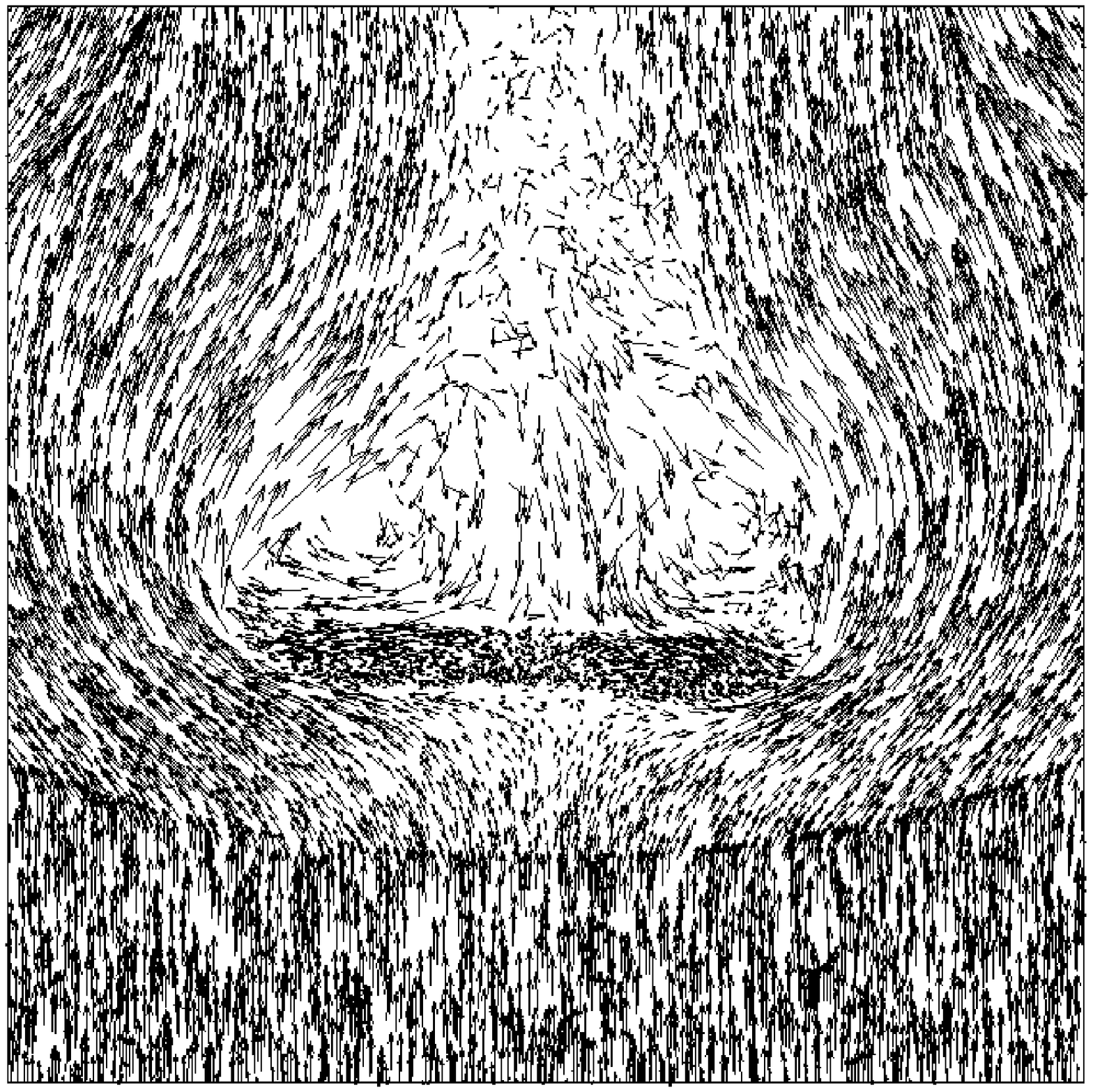}
\hspace{0.3truecm}
\includegraphics[angle=-90,scale=0.45]{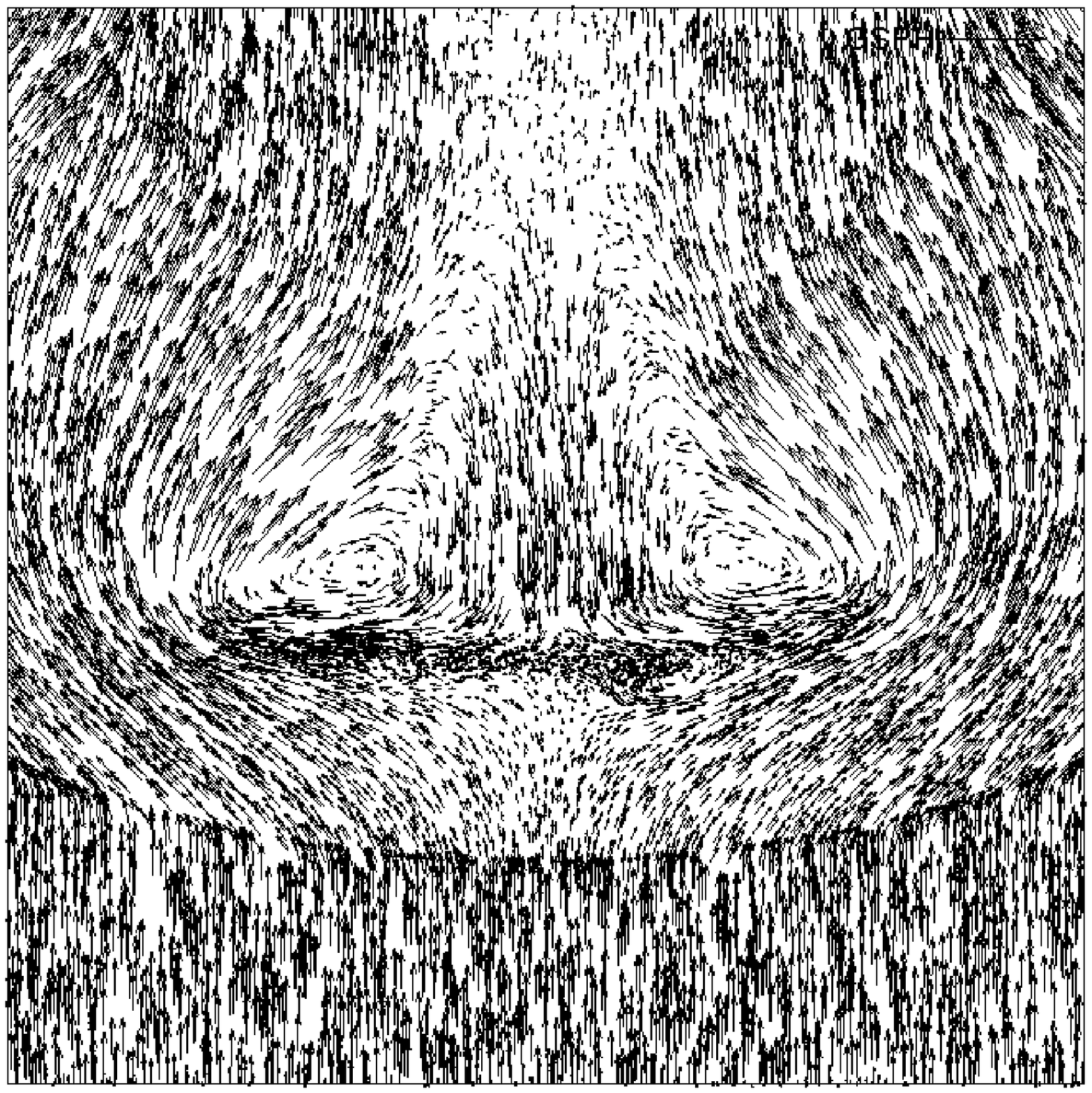}
}
\vspace{0.3truecm}
\hbox{
\includegraphics[angle=-90,scale=0.45]{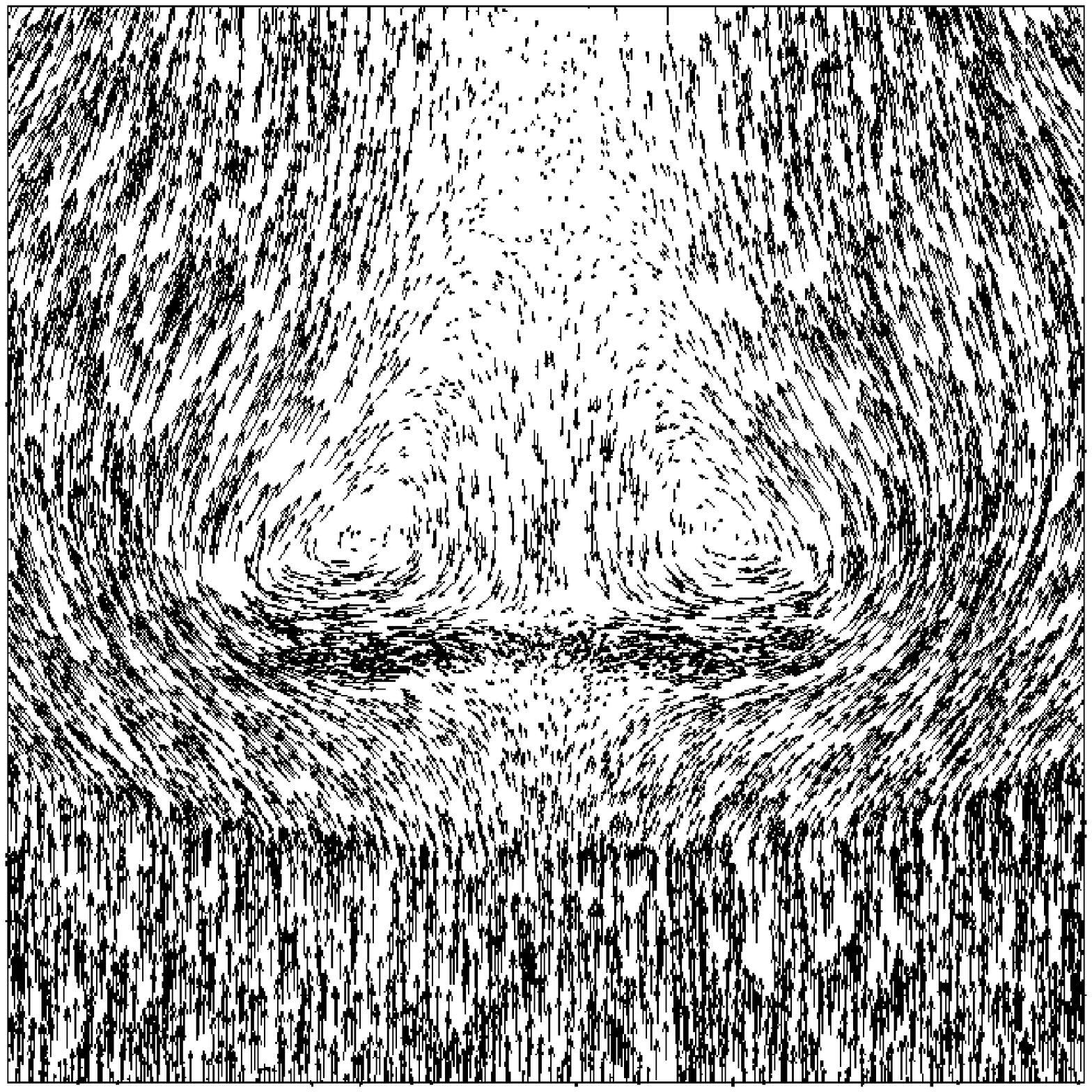}
\hspace{0.3truecm}
\includegraphics[angle=-90,scale=0.45]{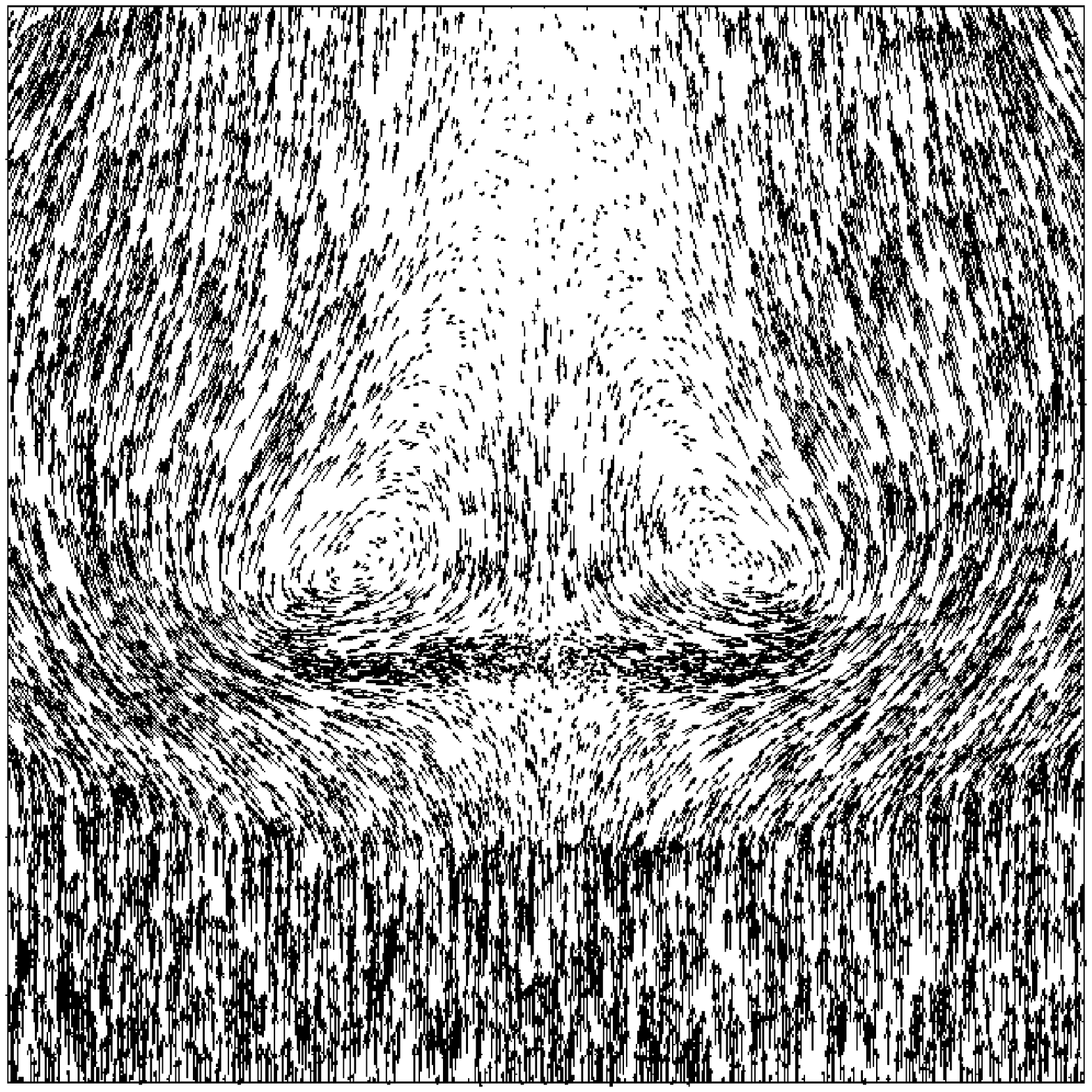}
}
}
\caption{Velocity field, at the time $t=4$, for the ``blob'' test, in
  the region surrounding the blob itself. We plot a slice near the
  center of the blob, with $900<x<100$ kpc, $400<y<1600$ kpc and
  $1500<z<3500$ kpc. We show the velocity field for \protect\GADGET
  (upper-left panel), \protect\GSPH (upper-right panel),
  \protect\GSPHCW\ (lower-left panel) and \protect\GSPHord\
  (lower-right panel). Each arrow shows the velocity of one gas
  particle; we undersampled the simulation by a factor 0.0025, for
  clarity. Velocities are computed in rest frame of the blob.}
\label{fi:vecvel}
\end{figure*}

\begin{figure*}
\vbox{
\hbox{
\includegraphics[angle=-90,scale=0.205]{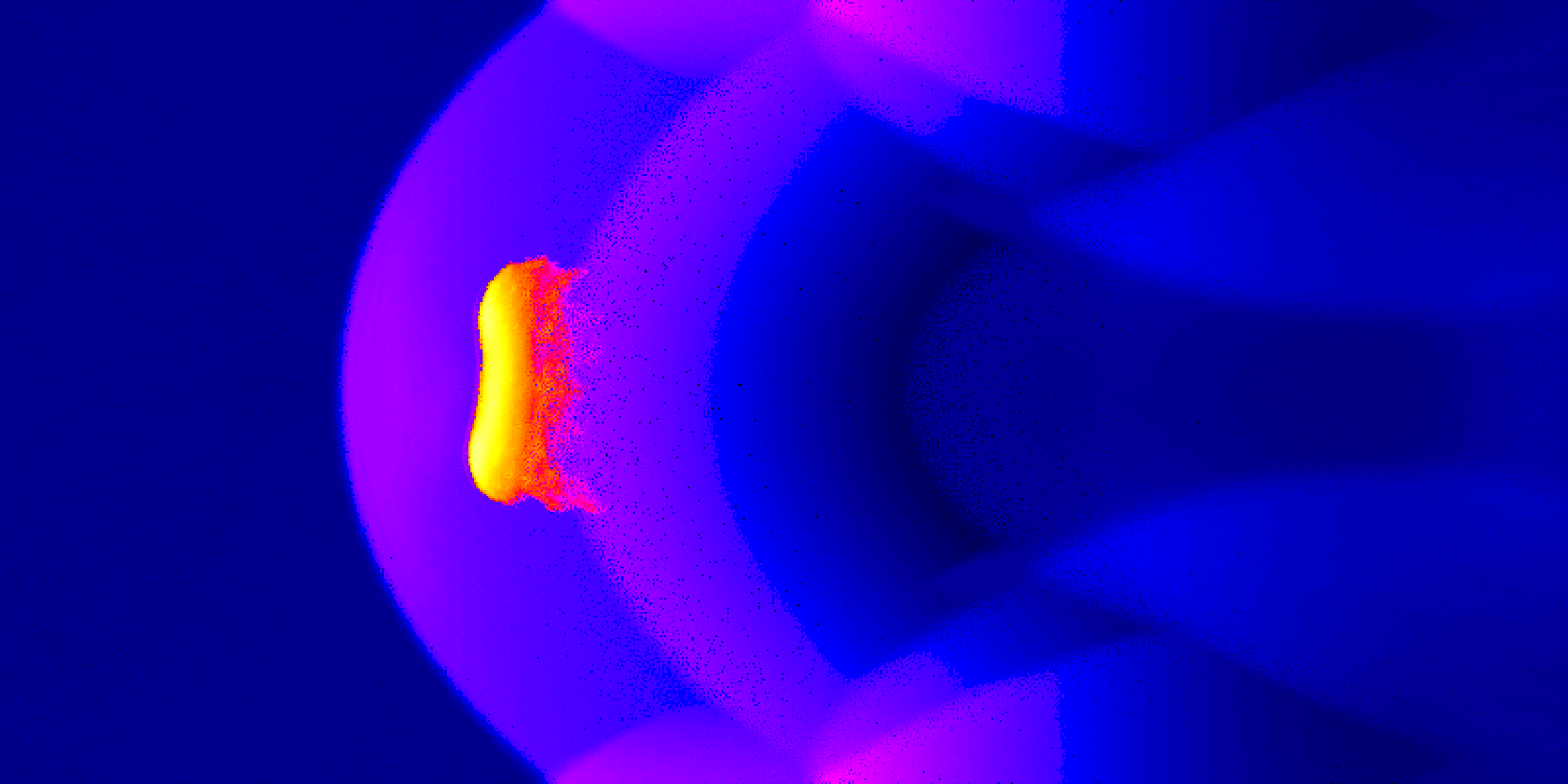}
\includegraphics[angle=-90,scale=0.205]{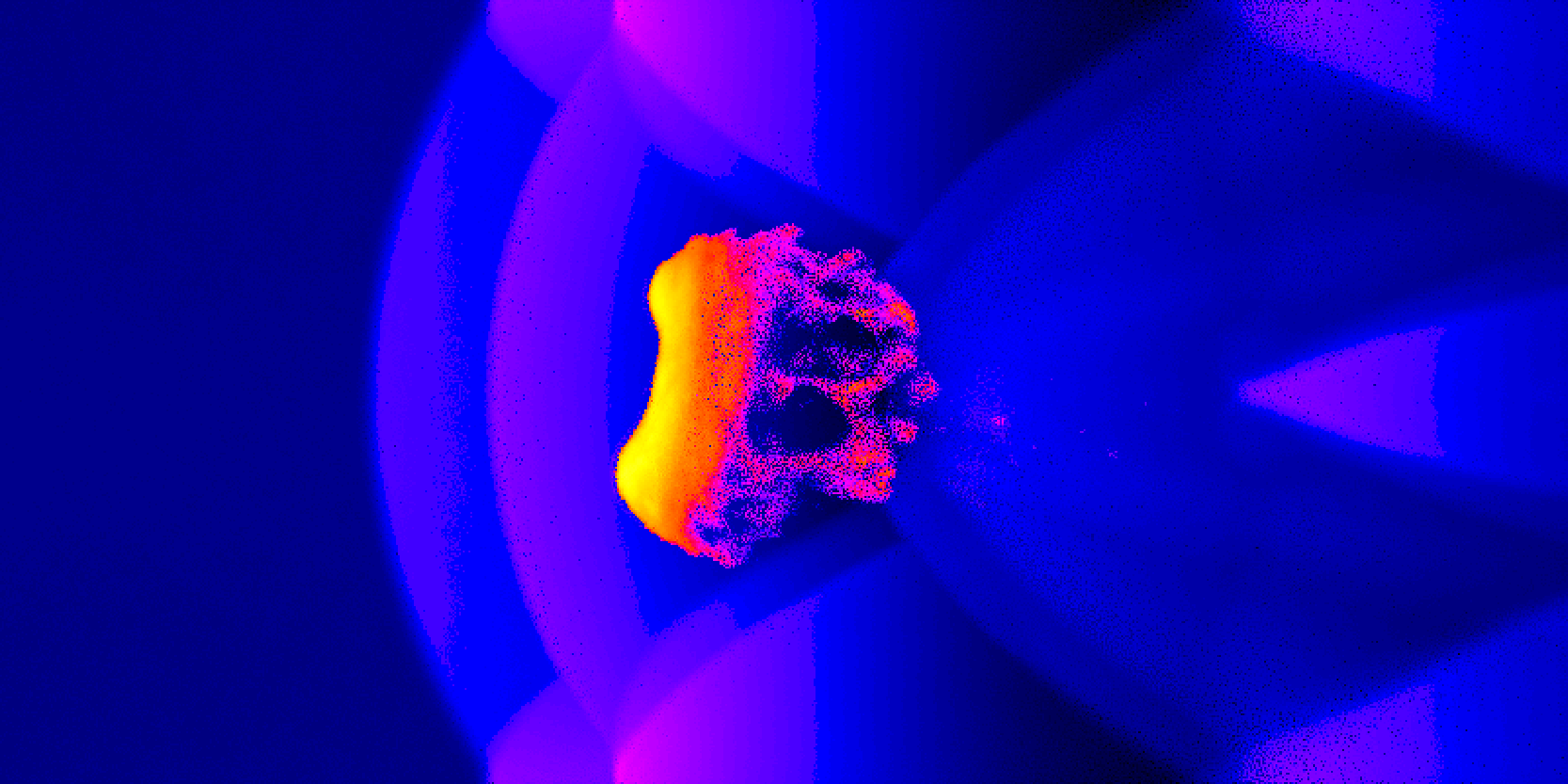}
\includegraphics[angle=-90,scale=0.205]{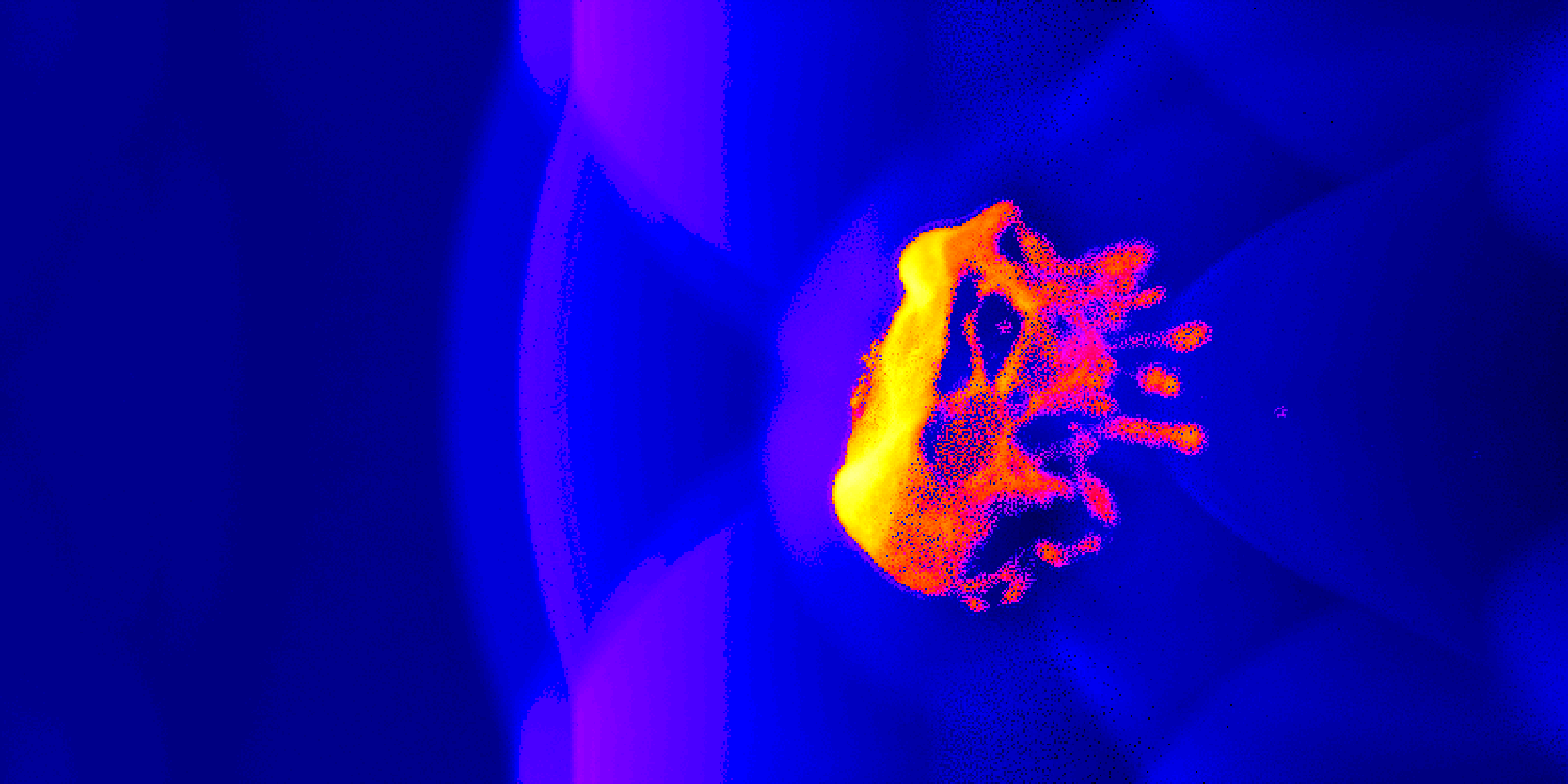}
}
\hbox{
\includegraphics[angle=-90,scale=0.205]{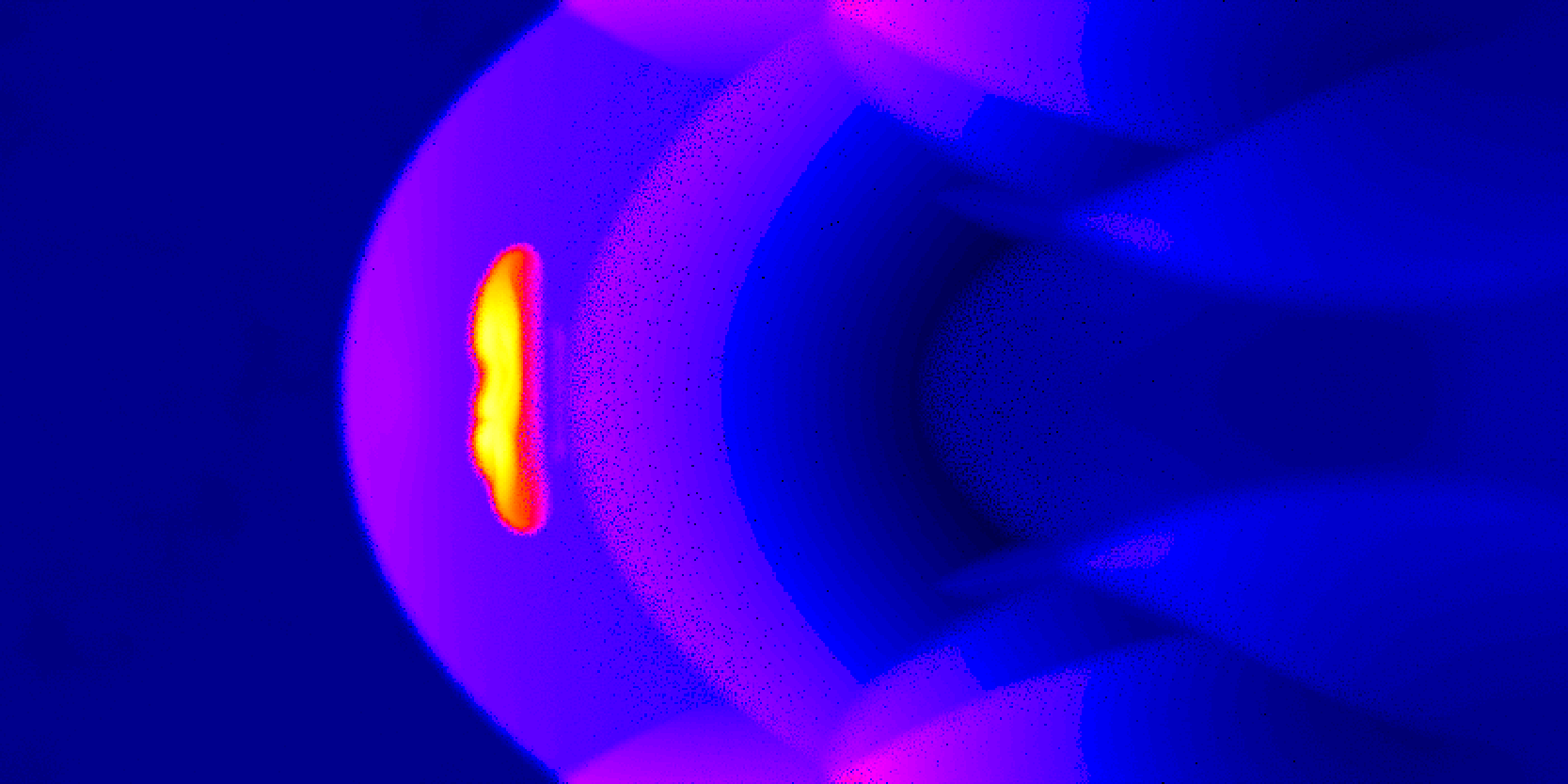}
\includegraphics[angle=-90,scale=0.205]{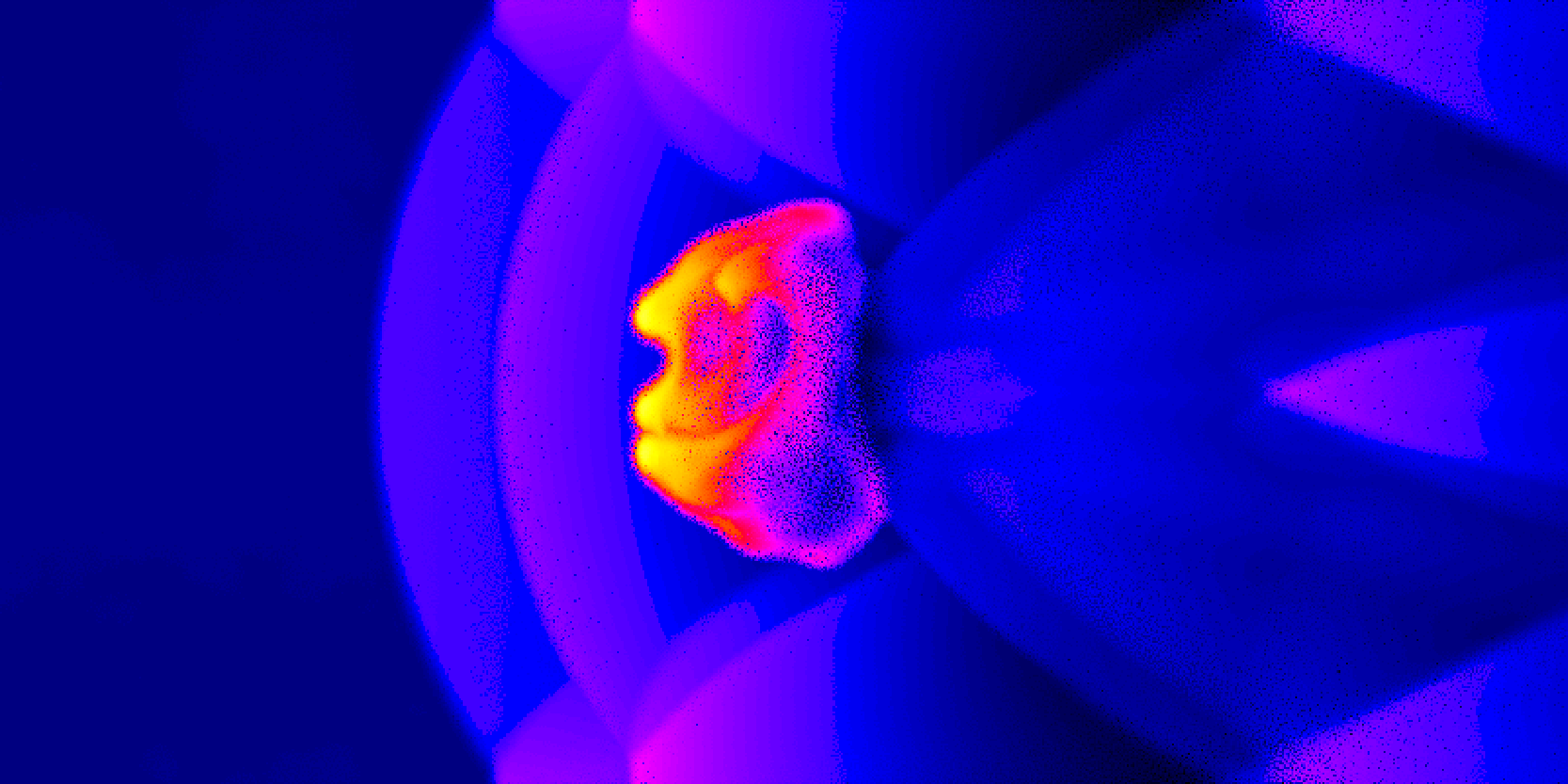}
\includegraphics[angle=-90,scale=0.205]{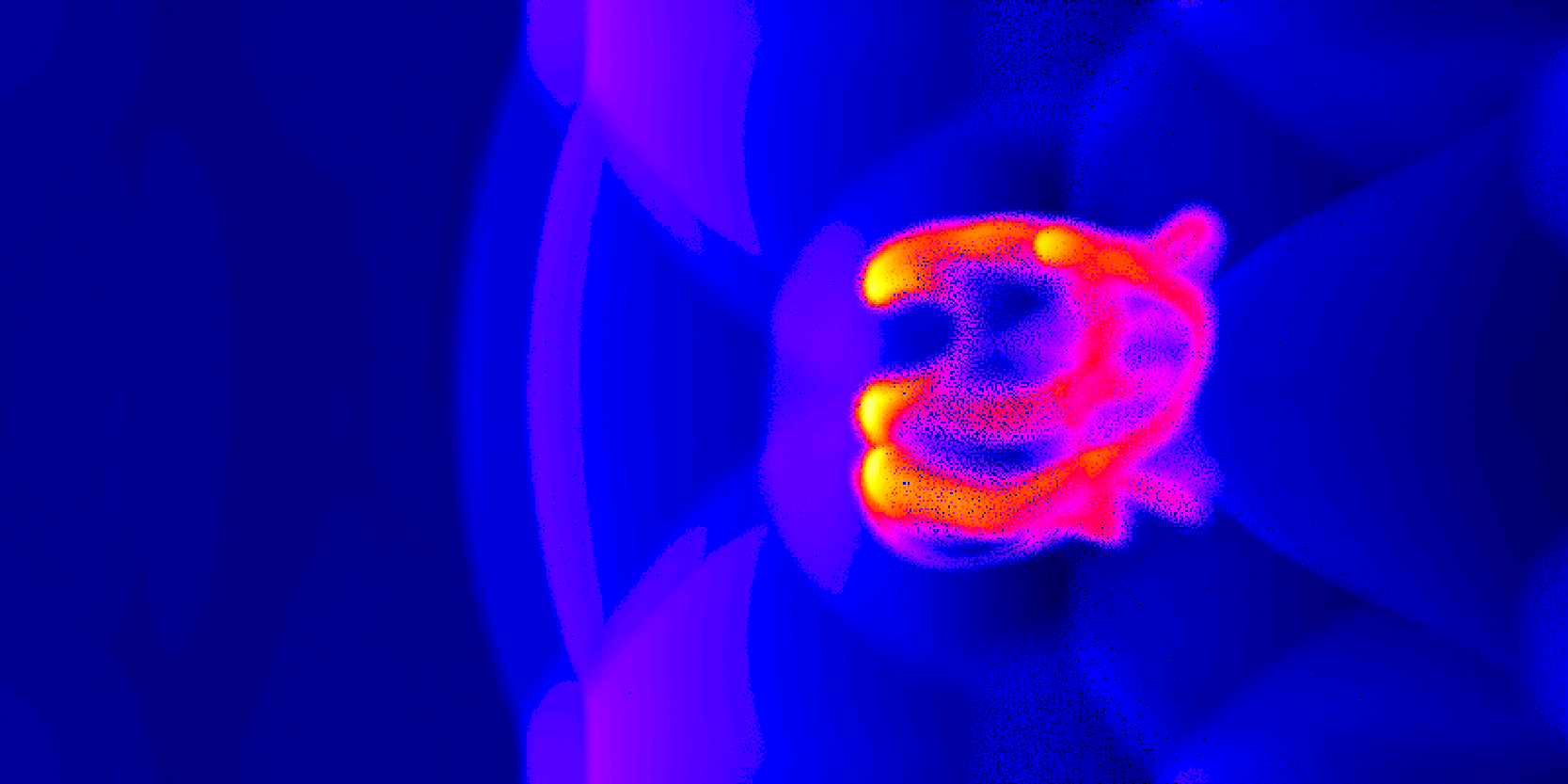}
}
\hbox{
\includegraphics[angle=-90,scale=0.285]{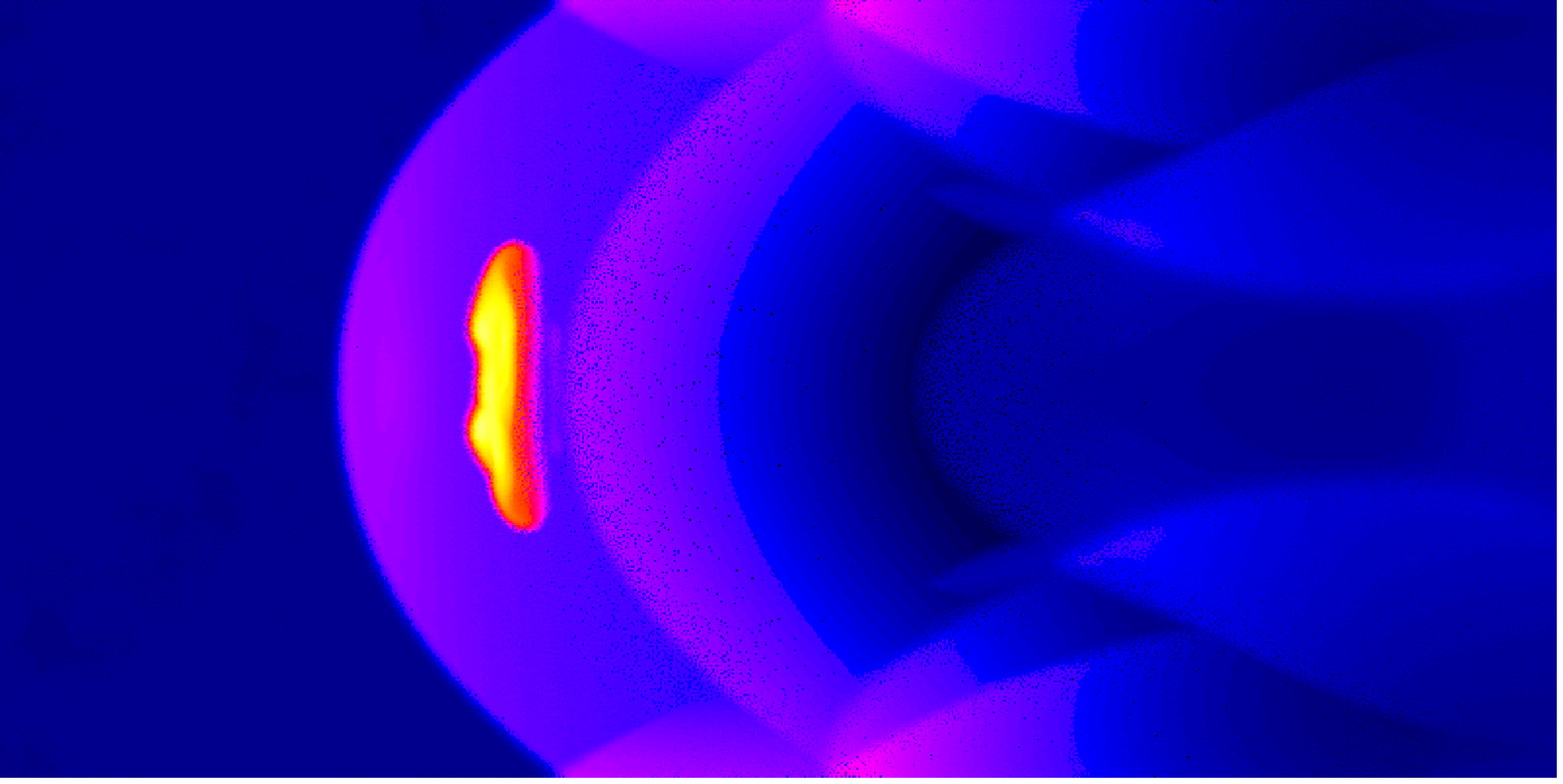}
\includegraphics[angle=-90,scale=0.285]{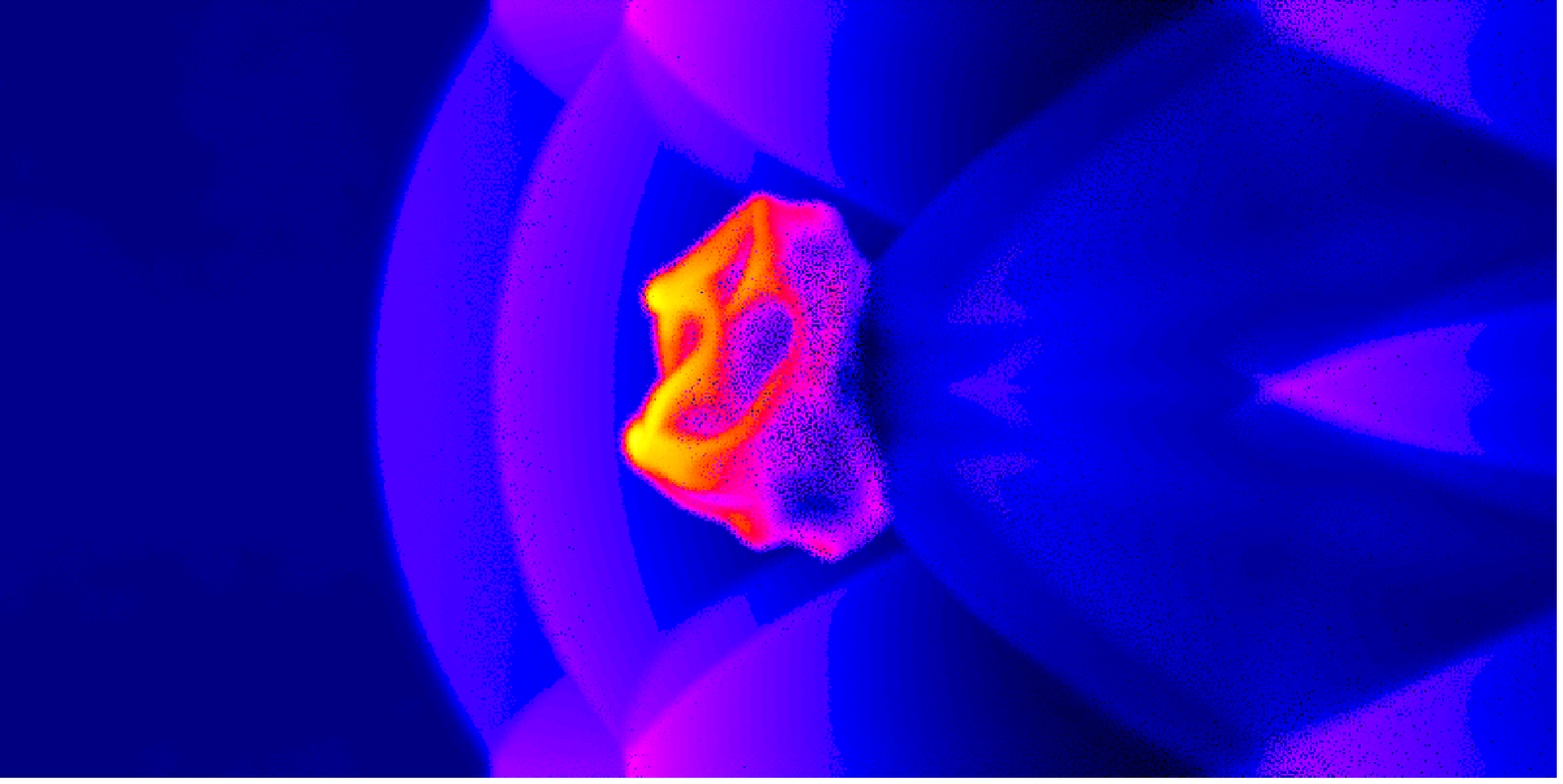}
\includegraphics[angle=-90,scale=0.285]{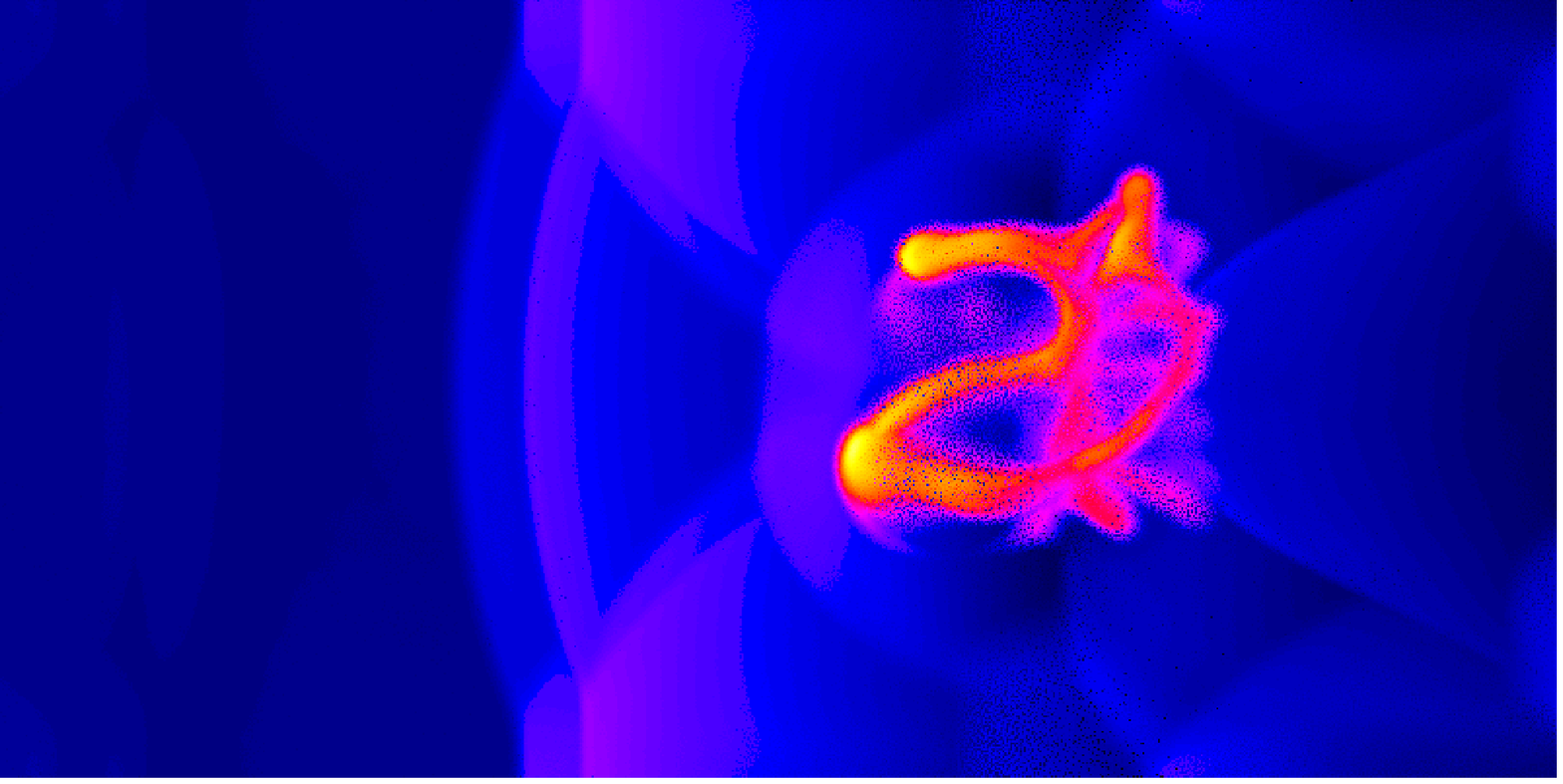}
}
}
\caption{Results for the ``blob'' test. We compare projected gas
  densities at $t=4$, $6$ and $8$ (left, central and right panels,
  respectively) for \protect\GADGET\ (upper panels), \protect\GSPHI\ (middle
  panels) and \protect\GSPH\ (bottom panels).}
\label{fi:blob}
\end{figure*}

Even if the total number of gas particles in our blob simulation
  is fairly high, force resolution, which is directly related to the
  mean interparticle separation, is lower than in the KH
  simulations. In fact, the mean interparticle separation is $14.7$
  kpc for the blob test, and $0.89$ kpc for the KH test. For this
  reason, the development of hydrodynamical instabilities is not
  clearly visible in Figure \ref{fi:blob}. In order to show the
  different behaviour of the different numerical schemes, we show in
  Figure \ref{fi:vecvel} the velocity field in a thin slice, centred
  on the blob, at the time $t=4$ when such instabilities begin to
  develop. The upper-left panel shows the field for GADGET, the
  upper-right one for \GSPH\ , the lower-left panel for \GSPHCW\ and
  the lower-right one for \GSPHord. Velocities are computed with
  respect to the rest-frame of the blob.

The difference between \GADGET\ and \GSPH\ is
  striking. Hydrodynamical instabilities create vortexes in the back
  of the blob in both cases. Quite clearly, the \GSPH\ scheme is by
  far most effective in resolving these vortexes and creating a
  downstream velocity flux. Also, the disruption of the blob front due
  to RT instabilities is marginally apparent in this figure. Note that
  the \GSPHCW\ scheme is also more effective than \GADGET\ in
  capturing the vortex structures. This is due to the absence of an
  artificial viscosity in such a scheme, which prevents the
  development of of vortex structure in the velocity field. However,
  the large diffusivity of \GSPHCW\ causes the suppression of the
  downstream flux. Finally, the behaviour of the \GSPHord\ scheme is
  intermediate between the other two GSPH schemes, since it is more
  diffusive than standard \GSPH\ but less so than \GSPHCW\ .

The difference of the \GADGET\ evolution with respect to the two GSPH
implementations shown in Fig. \ref{fi:blob} is even more apparent at
$t=6$. At this time the RT and RM instabilities appearing in the GPSH
simulations are further dissolved into filamentary and curl-like
structures, which are produced by the onset of KH instabilities. At
$t=8$ (right panels) the bulk of the gas initially contained in the
cold blob still forms a compact structure in the \GADGET\
simulation. On the contrary, in both GSPH simulations the blob is
basically dissolved at this time. The result for the \GSPHI\ case are
fully consistent with the two--dimensional blob test presented by
\cite{cha10}, who also used the I02 limiter. 

A comparison between the results obtained from the I02 and the
\cite{vanleer79} limiters show that they perform quite similarly in
this test. This suggests that differences between these two limiters
only becomes evident in higher resolution tests, such as the KH test
shown above.

In order to further verify the performances on the blob test of different
implementations of the GSPH scheme, we compare in Figure
\ref{fi:blob2} the results obtained at $t=8$ for the reference \GSPH\
scheme (upper left panel), for the \GSPHVLin\ version based on the
linear interpolation of the volume function (upper right panel), for
the \GSPHord\ version based on the first-order reconstruction for the
solution of the RP (bottom left panel), and on the \GSPHCW\ scheme
based on Eqs. ({\ref{eq:eom_gph}) and (\ref{eq:eneq_gph}). The result
for \GSPHVLin\ is rather similar to that of \GSPH, although the
latter develops a lower degree of filamentary structures and
instabilities. This result agrees with what shown in
Fig. \ref{fi:kh_gsph} for the KH test. On the other hand, the results
dramatically change if we use instead the first-order reconstruction
scheme of \GSPHord\ for the assignment of the thermodynamical
variables at the interface. In line with the KH test result, the
higher degree of diffusivity of this scheme dumps the development of
instabilities, thus preserving the structure of the blob. This result
highlights that, while using an accurate scheme of interpolation for
the volume function has a sizable effect, a much more dramatic change
is provided by properly choosing the reconstruction scheme for the RP
solution. Finally, the blob test confirms the result based on the KH
test on the incorrect description of the \GSPHCW\ scheme
in describing the development of instabilities.

In order to better quantify the different efficiency that
  different schemes have in describing the disruption of the blob, we
  show in Figure \ref{fi:massloss} the evolution of the blob mass loss
  for our various schemes. Following \citet{agertz07}, we define a gas
  particle to belong to the blob whenever its density is $\rho > 0.64
  \rho_{cl}$, being $\rho_{cl}$ the blob density as set in the initial
  conditions. As expected, \GADGET\ show the lowest degree of mass
  loss, a result which is in line with what shown by \cite{agertz07}
  and \cite{Hess10}. \GSPH\ , \GSPHVLin\ and \GSPHI\ have very similar
  mass loss rates, while \GSPHCW\ retains more mass at the end of our
  simulation. This can be understood in terms of a lesser ability of
  the latter scheme to capture the development of hydrodynamical
  instabilities, as shown in Figure \ref{fi:vecvel}. We also note that
  the strongest mass loss rate takes place for \GSPHord. Owing to the
  quite poor performance of this scheme in describing the development
  of the KH instability, we argue that numerical diffusion is in this
  case the main driver of the blob mass loss. This is also indicated
  by the different final shape of the blob (Figure
  \ref{fi:blob2}). The cloud is disrupted into several pieces in the
  \GSPH\ and \GSPHVLin\ schemes, while it retains its integrity in the
  \GSPHord\ and \GSPHCW\ ones. Therefore, significant mass loss of the
  blob can be obtained both through spurious numerical diffusion and
  through the development of genuine instabilities. However, only the
  latter are capable also to produce the disruption of the blob into
  several pieces.

\begin{figure}
\vbox{
\includegraphics[angle=-90,scale=0.35]{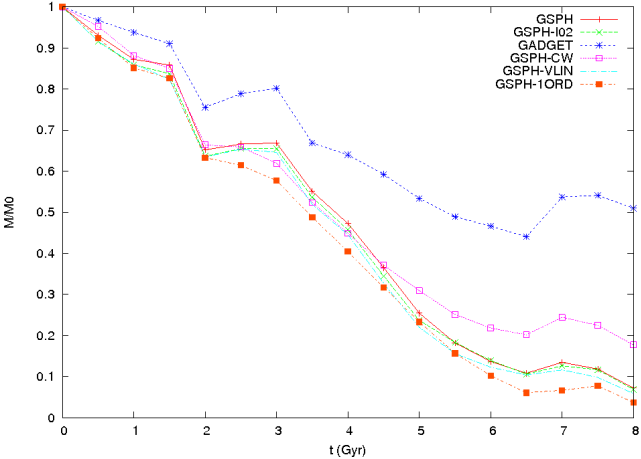}
}
\caption{Mass loss of the blob as a function of time for the different
  hydrodynamical schemes. We define a gas particle to be part of the
  blob if its density is $\rho > 0.64 \rho_{cl}$, with $\rho_{cl}$ the
  initial density of the blob. Red line with crosses refer to
  \protect\GSPH, green ones to \protect\GSPHI, blue ones to
  \protect\GADGET, magenta ones to \protect\GSPHCW, cyan ones to
  \protect\GSPHVLin\, and yellow ones to \protect\GSPHord.}
\label{fi:massloss}
\end{figure}

\begin{figure*}
\hspace{0.4truecm}
\vbox{
\hbox{
\includegraphics[angle=-90,scale=0.25]{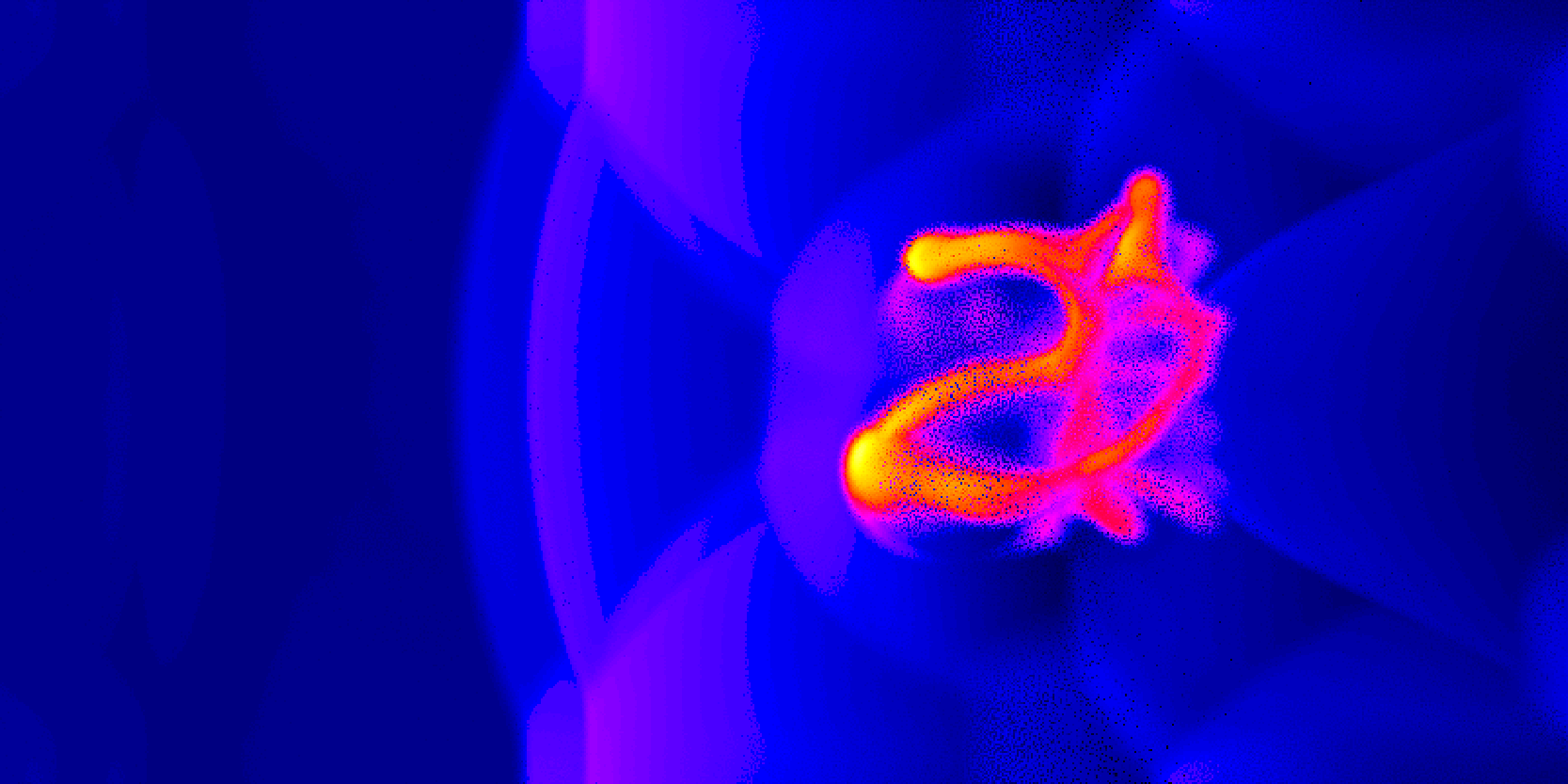}
\includegraphics[angle=-90,scale=0.25]{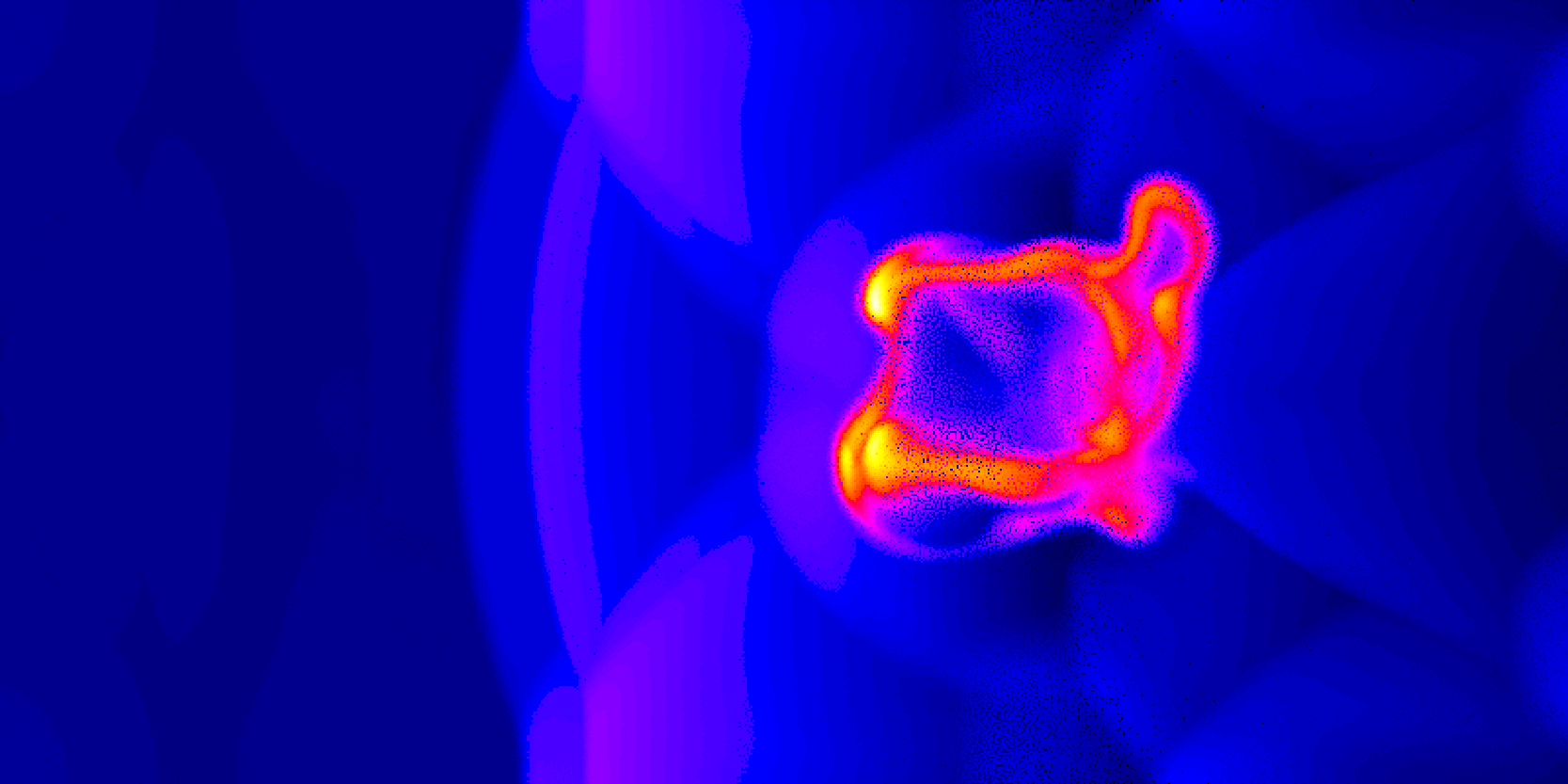}
}
\hbox{
\includegraphics[angle=-90,scale=0.25]{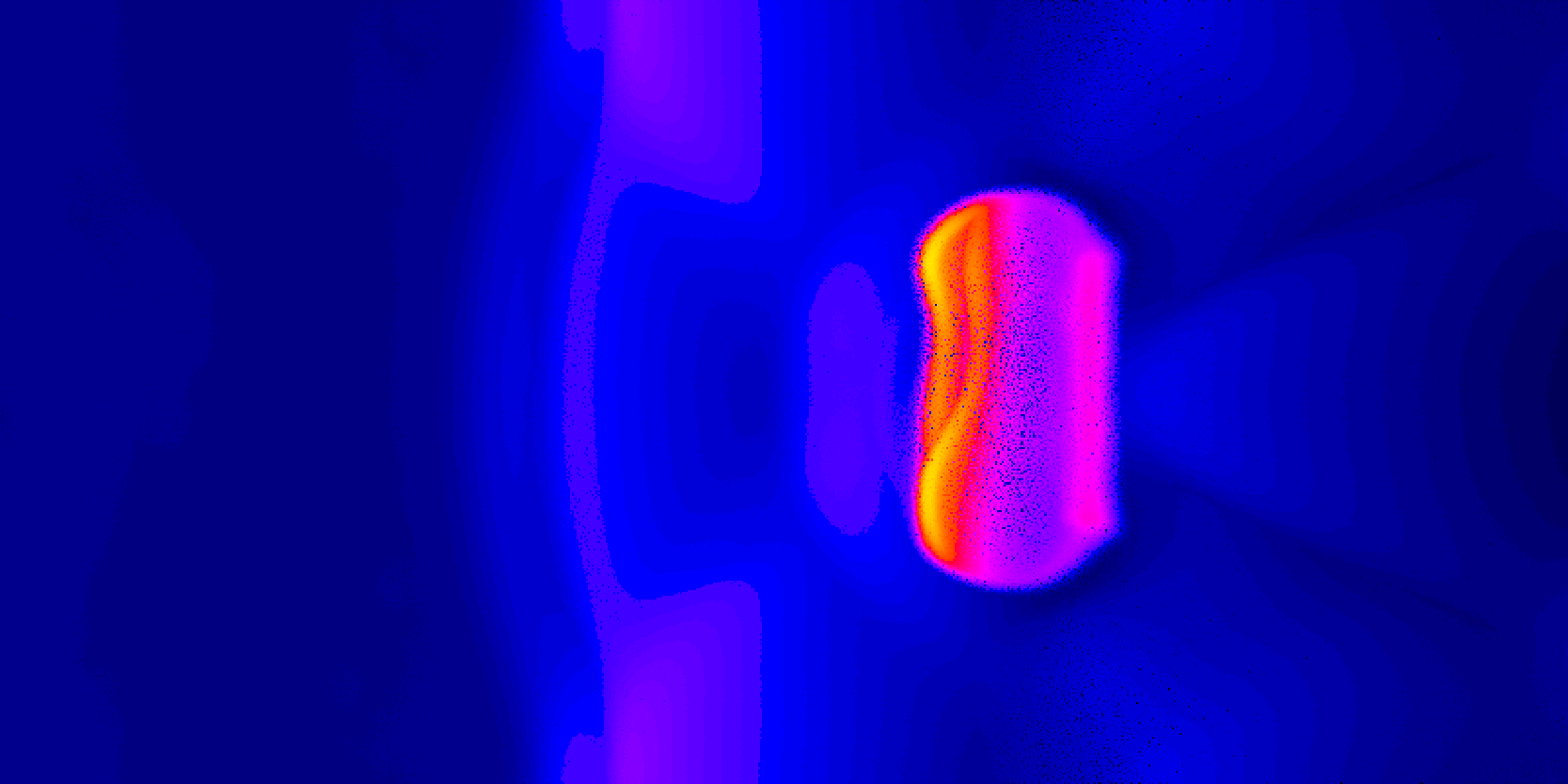}
\includegraphics[angle=-90,scale=0.25]{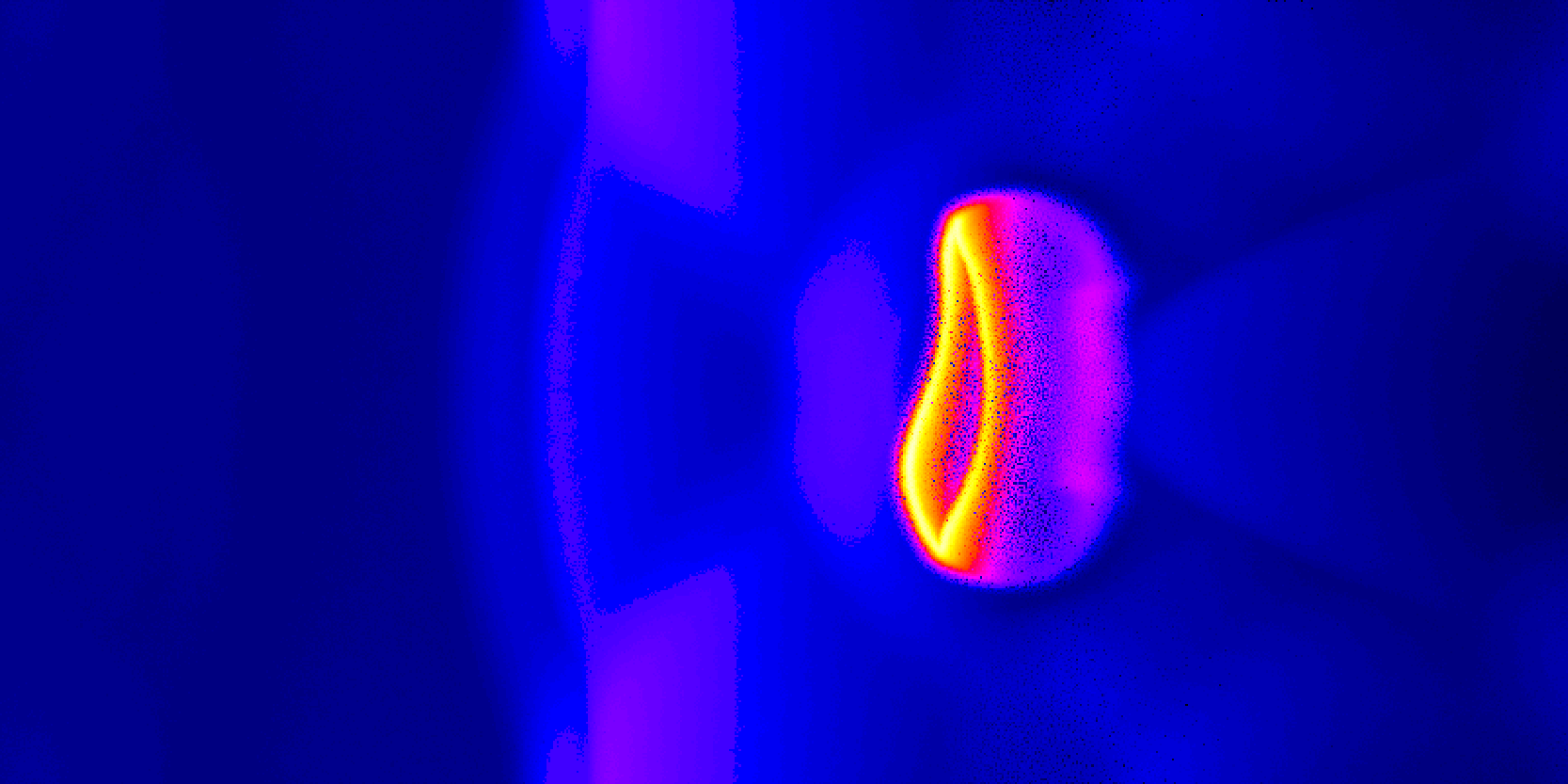}
}
}
\caption{Comparison of the results for the ``blob'' test at $t=8$ for
  different implementations of the GSPH scheme: \protect\GSPH\ (upper
  left), \protect\GSPHVLin\ (upper right), \protect\GSPHord\ (lower
  left), and \protect\GSPHCW\ (lower right).}
\label{fi:blob2}
\end{figure*}

\section{Conclusions}
In this paper we presented results of 3D standard hydrodynamical tests
for different implementations of the Godunov Smoothed Particle
Hydrodynamics (GSPH) within the \gadget\ code \citep{springel05}. The
conceptual difference between GSPH scheme and standard SPH scheme lies
in the fact that the former is based on explicitly convolving
momentum and energy equations with the interpolation kernel. The
resulting equations implemented in the simulation code are exact to
$O(h^2)$. Suitable expression for momentum and energy equations to be
implemented numerically are obtained by assuming a Gaussian shape for
the interpolating kernel. A natural way of implementing the equations
of \GSPH\ is by solving the Riemann Problem between each pair of
particles \citep[see][for a detailed discussion of
\GSPH]{inutsuka02}. Quite remarkably, solving the RP between each
particle pair brings the extra benefit that no artificial viscosity is
required to capture shocks, unlike in the standard SPH.

The different implementations of the \GSPH\ scheme, that we presented
in this paper (see Table 1), correspond to {\em (a)} using either
linear or cubic-spline interpolation for the volume function, which
provides the position of the interface where to solve the RP; {\em
  (b)} using either a first-order or a second-order reconstruction
scheme to assign thermodynamical variables at the interface in the
solution of the RP; {\em (c)} using different limiters to prevent
oscillations of interpolated quantities in the RP. Furthermore, we
also considered a variant of the \GSPH\ scheme, which is not based on
the convolution of the equations of fluido-dynamics, and is instead
essentially based on replacing pressure and velocity in the SPH
equations with the values obtained from the solution of the RP. The
performances of these different implementations to describe
discontinuities and development of gas-dynamical instabilities
have been assessed using a shock tube test \citep{sod78}, a shear-flow
test to follow Kelvin--Helmholtz (KH) instabilities and
the disruption of a cold blob moving in a hot atmosphere
\citep[e.g.,][]{agertz07}.

The results of our simulation tests can be summarised as follows.

\begin{description}
\item[(1)] As for the shock tube (see Fig. \ref{fi:sod}), we verified
  that \GSPH\ is able to correctly follow the development of the
  shock, despite the fact that it does not include artificial
  viscosity. Furthermore, \GSPH\ is also effective in removing the
  spurious ``pressure blip'' generated by standard SPH at the contact
  discontinuity, thanks to its capability to describe diffusion of
  entropy across the discontinuity (see Fig. \ref{fi:sod_dentr}). Quite
  interestingly, the best description of the discontinuity is provided
  by our reference \GSPH\ scheme, rather than by the more diffusive
  \GSPHCW\ scheme. This result highlights that including thermal
  diffusion is not enough in itself to provide a completely correct
  description of discontinuities. Indeed, the more accurate
  description of density gradients offered by the \GSPH, also
  significantly contribute to suppress spurious pressure forces at such
  discontinuities.
\item[(2)] Unlike standard SPH, our reference \GSPH\ scheme is quite
  effective in following the development of KH instabilities in the
  shear-flow test (see Fig. \ref{fi:kh}). We have verified that the
  accuracy in developing the curl structure of the instability is
  quite sensitive to the details of the implementation (see
  Fig. \ref{fi:kh_gsph}). For instance, using a first-order
  reconstruction dramatically degrades the \GSPH\ performance for this
  test. Also using a too small number of neighbours, a linear
  interpolation of the volume function and the limiter by
  \cite{inutsuka02}, instead of that by \cite{vanleer79}, also
  somewhat worsen, although to different degrees, the description of
  the KH instabilities.
\item[(3)] Similar results also hold for the ``blob test''. Also in
  this case, our standard \GSPH\ implementation follows the onset of
  Rayleigh-Taylor and Richtmyer-Meshkov (RT, RM) and KH instabilities.
  As a result, the blob is dissolved much more efficiently than in SPH
  simulations. However, the performance of \GSPH\ significantly worsen
  in case a first-order reconstruction scheme is used to assign
  variables at the interface. This highlights once again that
  diffusivity of the solution of the RP needs to be minimised to
  reliably follow the development of instabilities in \GSPH.
\end{description}

In this paper we focused on the comparison between different
hydrodynamical schemes, when applied to control test cases, without
analysing the behaviour of each of such schemes when resolution is
progressively increased (or degraded).  On the other hand, it is worth
reminding that a numerically diffusive scheme could converge to the
correct solution when applied to test cases with sufficiently high
resolution. For instance, \cite{Robertson10} showed this to be case
for the Galilean invariance in Eulerian codes, in the case of KH
instabilities.

In general our results agree with and extend those presented by
\cite{cha10} on two-dimensional KH and ``blob'' test of the GSPH. Our
analysis highlights the important role played by reconstruction at the
interface, by the choice of the limiter and by the interpolation order
for the volume function (see Appendix). The remarkable improvement
shown by \GSPH\ with respect to the standard SPH to describe contact
discontinuities and development of gas-dynamical instabilities makes
in principle this hydrodynamic scheme highly promising for applications
in computational astrophysics and cosmology.

As for its computational cost, it is worth pointing out that the need
of solving the RP between each pair of particles does not represent a
limiting factor. Clearly, the solution of the RP requires an iterative
procedure, which in principle could increase the computational
cost. This could be avoided using an approximate Riemann solver,
  which does not require an iterative procedure, as e.g. an
  Harten-Lax-van Leer-Contact (HLLC) solver.

However, the lack of artificial viscosity in the \GSPH\ makes
the Courant condition much less stringent in the shock regions than
for SPH, thereby leading to a more relaxed time-stepping. We verified
in our test that this  compensates the overhead associated to
the RP solution.

However, it is worth reminding that the \GSPH\ equations derived by
\cite{inutsuka02} \citep[see also][]{cha10} and used in our
implementation (Eqs. \ref{eq:eom_h} and \ref{eq:eneq_h}) hold only for
a Gaussian kernel. The subsequent request for a fairly large number of
neighbours and the $\sqrt 2$ multiplicative factor in front of the
kernel smoothing length in the above equations make the neighbour
search quite expensive. An implementation of \GSPH\ based instead on a
kernel with compact support, like the B-spline kernel, would clearly
be highly desirable to make the code more efficient for applications
involving large dynamic and temporal ranges.

\section*{Acknowledgements}
We are greatly indebted to Volker Springel for having provided us with
the non--public version of the GADGET-3 code and the initial
conditions of the shock tube test. We thank J. Read for providing us
with the initial conditions for the Kelvin--Helmholtz and ``blob''
tests. 
We also wish to thank an anonymous referee for useful suggestions
  that improved the presentation of the results. We acknowledge
useful and enlightening discussions with G. Bodo, K. Dolag,
R. Mignone. Simulations have been carried out on the IBM-SP6 machine
at CINECA (Bologna, Italy). This work has been partially supported by
the INFN PD-51 grant, by the PRIN-MIUR 2007 grant ``The Cosmic Cycle
of Baryons'', and by the PRIN-INAF 2009 grant ``Toward an Italian
Network for Computational Cosmology''.

\section*{Appendix. Volume interpolation}
\label{app_a}
In this Appendix we summarise for completeness the expressions for the
interpolating volume $V_{i,j}(h)$, which appears in the GSPH equations
of evolution (\ref{eq:eom_gsph}) and (\ref{eq:eneq_gsph}), in the case
of linear and of cubic spline interpolation. We will also provide the
corresponding expressions for the position of the interface at which
the Riemann problem between the $i$-th and the $j$-th particle is
solved. A full derivation of all such expressions is provided by I02.

The linearly-interpolated expression of the specific volume at the
coordinate $s$, along the axis joining the $i$-th and the $j$-th
particle, is 
\be
V(s)=\rho(s)^{-1}=C_{i,j}s+D_{i,j}
\label{eq:linvol}
\ee
where 
\ba
C_{i,j}&=&{V(\bx_i)-V(\bx_j)\over \Delta s_{i,j}}
\nonumber \\
D_{i,j}&=&{V(\bx_i)+V(\bx_j)\over 2}\,
\ea
We remind that we denote with $s_i$ and $s_j$ the
components of the $\bx_i$ and $\bx_j$ vectors along the $s$-axis, so
that $\Delta s_{ij}=s_i-s_j=|\bx_i-\bx_j|$.

Including the above expression for $\rho^{-1}$ into the integral
appearing on the {l.h.s.} of Eq.(\ref{eq:vij}), one obtains
\be
V_{i,j}^2\,=\,{1\over 4}h^2C_{i,j}^2 + D_{i,j}^2\,.
\label{eq:vij_lin}
\ee
In order to compute the position of the interface, let us define the
weighted--average $f_{i,j}^*$ of a generic function $f(\bx)$ through
the relation
\ba
&&\int {f(\bx)\over
  \rho^2(\bx)}W(\bx-\bx_i;h)W(\bx-\bx_j;h)d\bx=\nonumber \\
&&f_{i,j}^*\int {1\over
  \rho^2(\bx)}W(\bx-\bx_i;h)W(\bx-\bx_j;h)d\bx\,.
\ea
Using then the linear approximation for $f(s)=s(f_i-f_j)/\Delta
s_{i,j}$ and the above linear interpolation for $\rho(s)^{-1}$, one
obtains
\be
f_{i,j}^*={f_i-f_j\over \Delta s_{i,j}}s_{i,j}^*+{f_i+f_j\over 2}\,.
\ee
In the above equation the position of the interface
\be
s_{i,j}^*={h^2C_{i,j}D_{i,j}\over 2 V_{i,j}^2(h)}
\label{eq:sstar}
\ee
is defined as the position on the $s$-axis at which the
linearly-interpolated function $f$ takes the value $f_{i,j}^*$. 

The computation for a more accurate cubic--spline interpolation of the
volumes proceeds in a similar way. In this case, it is
\be
V(s)=\rho^{-1}(s)=A_{i,j}s^3+B_{i,j}s^2+C_{i,j}s+D_{i,j}
\ee
where the coefficients of the interpolating function are given by
Eqs.(61) of I02. Using again Eq.(\ref{eq:vij}) for the definition of
$V_{i,j}$, one obtains in this case
\ba
V_{i,j}^2&=&{15\over 64}h^6 A_{i,j}^2+{3\over
  16}h^4(A_{i,j}C_{i,j}+B_{i,j}^2)+\nonumber \\
&+&{1\over 4}h^2(2B_{i,j}D_{i,j}+C_{i,j}^2)+D_{i,j}^2
\ea
so that the expression for the position of the interface becomes
\ba
s_{i,j}^*&=&{1\over V_{i,j}^2(h)}]\biggl({15\over 32}h^6
A_{i,j}B_{i,j}+\nonumber \\ 
&+&{3\over 8}h^4(A_{i,j}D_{i,j}+B_{i,j}C_{i,j})+
{1\over 2}h^2 C_{i,j} D_{i,j}\biggr)\,.
\ea

\bibliographystyle{mn2e}
\bibliography{master_gsph}

\end{document}